\documentclass{aa}
\usepackage{graphics,lscape}
\begin{document}

\def\la{\mathrel{\hbox{\rlap{\hbox{\lower4pt\hbox{$\sim$}}}\hbox{$<$}}}}
\def\ga{\mathrel{\hbox{\rlap{\hbox{\lower4pt\hbox{$\sim$}}}\hbox{$>$}}}}
\def\sq{\hbox{\rlap{$\sqcap$}$\sqcup$}}
\def\arcmin{\hbox{$^\prime$}}
\def\arcsec{\hbox{$^{\prime\prime}$}}
\def\fd{\hbox{$.\!\!^{d}$}}
\def\fh{\hbox{$.\!\!^{h}$}}
\def\fm{\hbox{$.\!\!^{\rm m}$}}
\def\fs{\hbox{$.\!\!^{s}$}}
\def\fdg{\hbox{$.\!\!^\circ$}}
\def\farcm{\hbox{$.\mkern-4mu^\prime$}}
\def\farcs{\hbox{$.\!\!^{\prime\prime}$}}

\newcommand{\etal}{{et al.}\,}      % et al. in italics
\newcommand{\eg}{{e.g.},\ }         % e.g. in italics
\newcommand{\ie}{{i.e.},\ }         % i.e. in italics
\newcommand{\cf}{{\it cf.},\ }          % c.f. in italics

\def\deg{{^\circ}}

%\thesaurus{03(01.01.1; % Catalogues 
%              03.04.1; % Surveys
%              07.09.1; % ISM: dust, extinction
%              09.13.1; % Galaxies: fundamental parameters
%              20.01.2) % Cosmology: large-scale structure of Universe
%                      }

\title{A catalogue of galaxies behind the southern Milky Way.\thanks{The two optical catalogues and their respective
listings of IRAS cross-identifications are available in electronic format at the CDS
via anonymous ftp to cdsarc.u-strasbg.fr (130.79.128.5) or via
http://cdsweb.u-strasbg.fr/Abstract.html}}
\subtitle{II. The Crux and Great Attractor regions ($l\approx$ 289$\degr$ to 338$\degr$)}

\author{Patrick A. Woudt\inst{1} \and  Ren\'ee C. Kraan-Korteweg\inst{2}}

\offprints{Patrick A. Woudt,
\email{pwoudt@artemisia.ast.uct.ac.za}}

\institute{Dept.~of Astronomy, University of Cape Town, Private Bag, Rondebosch 7700, South Africa
\and
Depto.~de Astronom\'\i{a}, Universidad de Guanajuato, Apartado Postal 144, 
Guanajuato, GTO 36000, Mexico}

\date{Received date;accepted date}

\titlerunning{Galaxies behind the southern Milky Way. II.}
\authorrunning{P.A.~Woudt \& R.C.~Kraan-Korteweg}

\abstract{In this second paper of the catalogue series of galaxies behind the 
southern Milky Way, we report on the deep optical galaxy search in the Crux region ($289\deg \le \ell 
\le 318\deg$ and $-10\deg \le b \le 10\deg$) and the Great Attractor region ($316\deg \le \ell \le 338\deg$ and 
$-10\deg \le b \le 10\deg$).
The galaxy catalogues are presented, a brief description of the galaxy search given, 
as well as a discussion on the distribution and characteristics of 
the uncovered galaxies.
A total of 8182 galaxies with major diameters $D \ga 0\farcm2$ were 
identified in this $\sim$850 square degree area: 3759 galaxies in the Crux region
and 4423 galaxies in the Great Attractor region. Of the 8182 galaxies, 229 
(2.8\%) were catalogued before in the optical (3 in radio) and 251 galaxies have a reliable (159),
or likely (92) cross-identification in the IRAS Point Source Catalogue (3.1\%).
A number of prominent overdensities and filaments of galaxies are identified. 
They are not correlated with the Galactic foreground extinction 
and hence indicative of extragalactic large-scale structures.
Redshifts obtained at the South African Astronomical Observatory (SAAO) for 518 of the 
newly catalogued galaxies in the Crux and Great Attractor regions (Fairall \etal 1998; Woudt
\etal 1999) confirm distinct voids and clusters in the area here surveyed.
With this optical galaxy search, we have reduced the
width of the optical `Zone of Avoidance' for galaxies with {\sl extinction-corrected} 
diameters larger than 1.3 arcmin from extinction levels $A_B \ge 1\fm0$ to 
$A_B \ge 3\fm0$: the remaining optical 
Zone of Avoidance is now limited by $| b |  \la 3^\circ$ (see Fig.~\ref{cruxf1new}).
}

\maketitle

\section{The Great Attractor and the Zone of Avoidance}

Dust and stars in the plane of the Milky Way obscure $\sim$20\% of the optical extragalactic sky
and 10\% of the IRAS extragalactic sky. As a result, existing 
optical galaxy catalogues are severely incomplete close to the Galactic Equator leading to a
`Zone of Avoidance' (ZOA) in the distribution of galaxies. 
For example, the main optical galaxy catalogue of the southern 
sky (Lauberts 1982) is complete for galaxies with an observed diameter $D \ge 1$\farcm3 
(Hudson and Lynden-Bell 1991) down to extinction-levels of $A_B \le 1^{\rm m}$ (see Fig.~1 
of Kraan-Korteweg \& Lahav 2000). At higher extinction-levels, galaxies with an intrinsic diameter of
1\farcm3 fail to meet the selection criteria (Lauberts' (1982) selection criterion
is $D_{obs} \ge 1'$) and only the intrinsically largest and brightest 
galaxies are detected near the Galactic Plane.

This incompleteness limits our understanding of the origin of the peculiar motion of the
Local Group with respect to the Cosmic Microwave Background, and the origin of velocity
flow fields in the local Universe.

In our previous Zone of Avoidance (ZOA) catalogue paper (Kraan-Korteweg 2000a, hereafter Paper I), a 
detailed motivation for our deep optical galaxy search behind the southern Milky Way was given. 
The main arguments for embarking on a survey of this nature are briefly reiterated here.

\begin{figure*}
\centering
 \resizebox{\hsize}{!}{\includegraphics{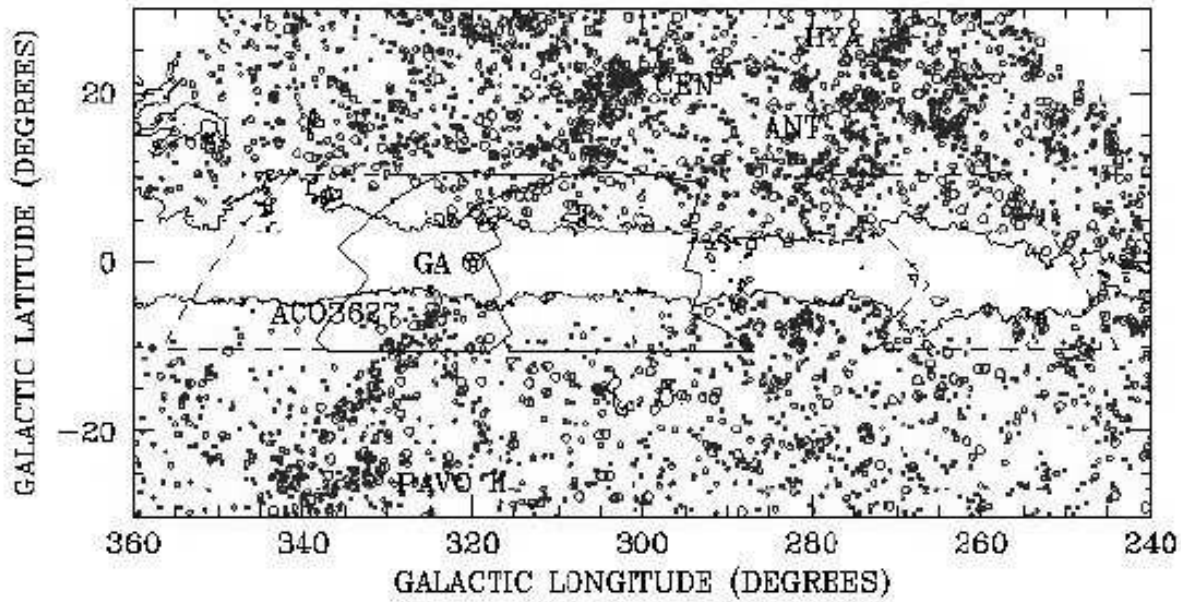}}
 \caption{The distribution in Galactic coordinates of Lauberts (1982) galaxies
 with extinction-corrected diameters $D^0 \ge 1$\farcm3 and $A_B \le 3^{\rm m}$, supplemented with
 galaxies from the deep galaxy search in the Hydra/Antlia region (Kraan-Korteweg 2000a) with the same limits. 
 The contour is a line of equal Galactic foreground extinction, taken from the Galactic 
 reddening maps of Schlegel \etal (1998), and corresponds to $A_B = 3^{\rm m}$.
 The galaxies are diameter-coded: the galaxies
 with 1\farcm$3 \le D^0 \le 2'$ are displayed as small circles, the galaxies
 with $2' \le D^0 \le 3'$ as medium-sized circles and the galaxies with $D^0 \ge 3'$
 as large circles. The thick solid line outlines the Crux (right) and Great Attractor (left) search areas.
 The dotted line marks the other search areas: 
 the Scorpius region (adjacent to the Great Attractor region) and the Vela region (to the right). 
 The Centaurus, Pavo II, Centaurus-Crux and ACO3627 clusters are 
 labelled, as is the peak of the reconstructed mass density 
 field associated with the Great Attractor. }
 \label{cruxf1}
\end{figure*}

\begin{itemize}
\item[$\bullet$]{To improve the determination of the optical galaxy density field across the sky through
the reduction of the ZOA. By directly observing
the galaxy distribution in the ZOA -- contrary to inferring the galaxy density field
from the velocity flow field (\eg Kolatt \etal 1995), or interpolating the galaxy density
field outside the ZOA into the ZOA (Yahil \etal 1991) -- and comparing this observed distribution
with the velocity flow field, fundamental cosmological parameters (such as $\beta = \Omega^{0.6}/b$,
where $b$ is the linear biasing parameter) can be derived (Strauss \& Willick 1995).}
\end{itemize} 
In Paper I we have already noted that the deep optical galaxy searches lead to a {\sl complete} optical
galaxy distribution for galaxies with extinction-corrected diameters $D^0 \ge 1\farcm3$ for $A_B \le 3^{\rm m}$, resulting
in a reduction of the optical ZOA of over 50\% (see also Fig.~4 of Kraan-Korteweg \& Lahav 2000).

\begin{itemize}
\item[$\bullet$]{To unveil the full extent of the Great Attractor. The Great Attractor (GA) is
seen primarily in the peculiar velocity field of galaxies in the local Universe (Dressler \etal
1987; Lynden-Bell \etal 1988; Kolatt \etal 1995; Tonry \etal 2000).
There is no doubt that this overdensity exists.
There is, however, still some ambiguity about the true nature
and extent of the Great Attractor (e.g.,~Staveley-Smith \etal 2000). This is primarily caused by its unfortunate location right
behind the southern Milky Way at ($\ell, b, v$) $\approx$ ($320\deg, 0\deg, 4000$ km s$^{-1}$)
(Kolatt \etal 1995). It is very likely that dust and stars in our Galaxy have greatly diminished
the optical appearance of the GA, and that a significant fraction (of the mass) of the 
GA overdensity lies behind the Milky Way.}
\end{itemize}

Figure~\ref{cruxf1} shows the complete diameter-limited southern sky distribution
of galaxies, down to a diameter-limit of $D^0 \ge 1$\farcm3. Only galaxies for which the 
foreground extinction is less or equal than $A_B \le 3^{\rm m}$ are shown. The diameters have been
corrected for the diminishing effects of the Galactic foreground extinction (Cameron 1990).
For the extinction correction we have used the Galactic reddening maps of
Schlegel \etal (1998). The results of our survey in the Hydra/Antlia region (Paper I) have been
included in this graph. 

The regions under investigation in this paper, the Crux region ($\ell \approx 289^\circ - 318^\circ$)
and the Great Attractor region ($\ell \approx 316^\circ - 338^\circ$), are demarcated by the thick
solid line. They lie in between the Hydra/Antlia region (Paper I) and the Scorpius region. 
The Crux and Great Attractor regions are of particular interest due to their
proximity to the Great Attractor. If a large fraction of the mass associated with the 
Great Attractor has remained hidden behind the Milky Way, this survey 
should reveal that.

In Sect.~2 we briefly describe the galaxy search. The catalogues are presented in Sect.~3, 
and in Sect.~4 we discuss the characteristics
of the magnitudes and diameters of the galaxies in our survey. In Sec.~5 we discuss the Galactic foreground extinction
and in Sect.~6 we assess the completeness of our survey.
The performance of the IRAS Point Source Catalogue for studies of large-scale structures at low Galactic latitudes
is discussed in Sect.~7. In Sect.~8, we assess the impact of our
survey on the current understanding of the Great Attractor.

\section{The galaxy search}

The search area is divided into five separate regions as illustrated in Fig.~\ref{cruxf1}: the 
`Hydra/Antlia region' (Paper I), the `Crux region' 
(this paper), the `Great Attractor region' (this paper), the `Scorpius region' 
(Kraan-Korteweg \& Fairall, in prep.) and the `Vela region' (Kraan-Korteweg \& Salem, in prep.).
 
\begin{figure*}
%\sidecaption
\resizebox{\hsize}{!}{\includegraphics{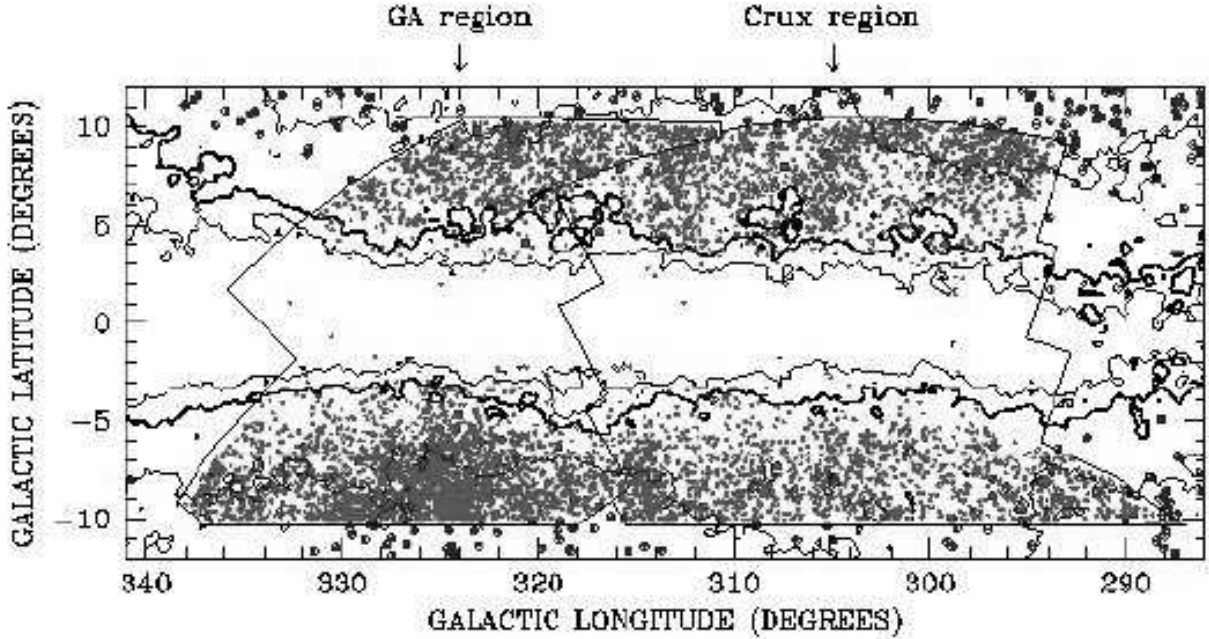}}
    \caption{The distribution in Galactic coordinates of galaxies in the Crux and Great Attractor regions. 
The search areas are marked. The 8182 unveiled galaxy candidates 
(with $D\ga 0$\farcm2) are shown as small dots. The encircled dots are Lauberts (1982) galaxies. The contours are lines of equal Galactic 
foreground extinction, taken from the Galactic reddening maps of Schlegel \etal (1998). 
The contours correspond to $A_B = 1\fm0$, $3\fm0$ (thick line), and $5\fm0$.}
     \label{cruxgaf2}
\end{figure*}

An extensive description of the galaxy search is given in Paper I. 
The IIIaJ film copies of the SRC sky survey have been examined systematically 
by eye in a darkened room, using a proto-type blinking machine (on
semi-permanent loan from the Astronomisches Institut der 
Universit{\"a}t Basel) with a 50 times magnification. 
An area of 3$\farcm$5 x 4$\farcm$0 is projected on a screen.

The selection of the IIIaJ film copies of the SRC sky survey as our plate material was made
after careful tests (Paper I). Even though the effects of the Galactic foreground 
extinction are stronger in the blue, the hypersensitized and fine grained emulsion of the 
IIIaJ films go deeper and show more resolution, compared to their red counterparts.

We imposed a diameter limit of $D \ga 0.2$ arcminutes 
for our search. For every galaxy we recorded the major and minor diameter.
In addition, we made an estimate of the average surface brightness and the morphological 
type of the galaxy. The magnitude of each galaxy was derived using the diameters and the estimated
average surface brightness. Finally, the positions of all the galaxies were measured 
with the Optronics machine at ESO in Garching. 

The Crux and Great Attractor regions together cover approximately 850 square degrees, ranging in 
Galactic longitude from $\ell \approx 289\deg$ to $\ell \approx 338\deg$ and limited
at $|b| \approx 10\deg$. It contains 
37 fields of the ESO/SRC sky survey, namely
F62--67, F94--100, F130--138, F171--180 and F221--225. These film copies were systematically 
examined by eye by the first author (PAW) with the magnifying viewer.

In total, 8182 galaxy candidates have been identified: 3759 in the Crux region
and 4423 in the Great Attractor region, respectively.

\subsection{The distribution of galaxies in the Crux region}

The resulting distribution of the 3759 galaxies in the Crux region is shown in Galactic coordinates in 
Fig.~\ref{cruxgaf2}. The search area is outlined. Next to the expected dependence
of the number density on Galactic latitude, strong variations with
Galactic longitude are evident. Folding the galaxy distribution with the Galactic reddening maps
of Schlegel \etal (1998), indicates that these density fluctuations are indeed extragalactic of
origin, even though dark clouds {\sl do} cause holes in the 
galaxy distribution (see for instance the galaxy distribution 
at ($\ell, b) \approx (301\deg, -9\deg$). 
Most noticable in this respect are the dark clouds around $\ell \approx 316\deg - 318\deg$ and
$b \approx +5\deg - +6\deg$. 
However, the observed overdensities and filamentary structures in the galaxy distribution 
can, overall, not be explained by extreme transparent regions.

Three filamentary structures are seen in the Crux region. Two of them are located 
above the Galactic Plane (GP), around $\ell \approx 305\deg$ and $\ell \approx 313\deg$. These filaments are a likely 
continuation of the Centaurus Wall (Fairall \etal 1998) into the ZOA.
The filament around $\ell \approx 305\deg$ could, however, also be a part of a larger 
overdensity extending to lower Galactic longitudes, incorporating the overdensity at 
($\ell, b) \approx (297\deg, +9\deg$), since dark clouds (Feitzinger and St\"{u}we 1984) 
possibly obscure part of this overdensity. One filament below the GP at $\ell \approx 315\deg$ has, 
however, no visible counterpart in the galaxy distribution beyond the ZOA. Given the relative 
small diameters of the galaxies in this filament, it most likely is at larger distances
and not connected with the Centaurus--Pavo connection, \ie the Centaurus Wall.

Number density contours also reveal an overdensity around
$(\ell, b) \approx (305.5\deg, +5.5\deg)$ which could mark a possible cluster of galaxies.
Redshifts taken at the South African Astronomical Observatory 
seem to confirm this cluster, given the small finger-of-god seen
at this position (see Fig.~6 of Fairall \etal 1998). This low-mass cluster (the Centaurus-Crux cluster) is part of the 
Great Attractor overdensity at a mean redshift-distance of 6214 km s$^{-1}$, albeit
on the far side of the Great Attractor.

\subsection{The distribution of galaxies in the Great Attractor region}

\begin{figure}
%\sidecaption
\resizebox{\hsize}{!}{\includegraphics{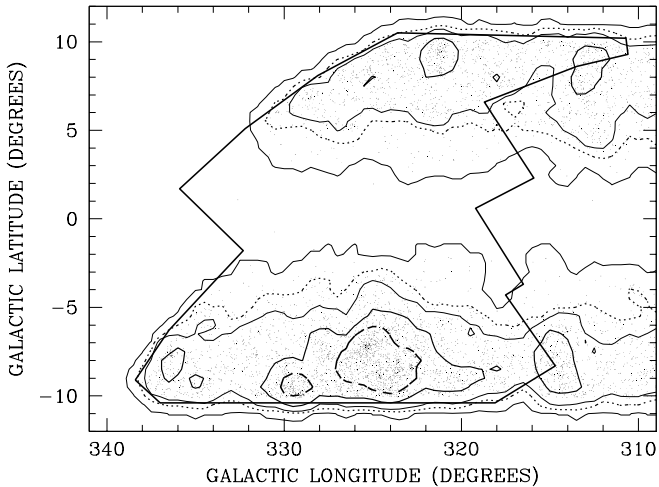}}
    \caption{The galaxy density distribution in the GA region. The contours
mark 0.5, 5 (dotted line), 10, 25 (thick solid line) and 50 (thick dashed line) galaxies per square degree, respectively.}
     \label{gaf3}
\end{figure}

Also shown in Fig.~\ref{cruxgaf2} is the distribution in Galactic coordinates of the 4423 galaxies 
that were found in the GA region. In addition, we have displayed in Fig.~\ref{gaf3} the
galaxy density contours in the GA region. Examination of both figures reveals 
three distinct overdensities: one above the Galactic Plane at $(\ell, b) = (321\deg,
+9\deg)$, and two at negative Galactic latitudes, at $(\ell, b) = (329\deg, -9\deg)$
and $(\ell, b) \approx (325\deg, -7\deg)$, respectively.
The latter is by far the most prominent overdensity of galaxies 
in the southern Zone of Avoidance (including the surveyed Hydra/Antlia and Crux region).
This overdensity is a factor $f = 8-10$ more dense compared to regions at similar
Galactic latitude. It is associated with ACO 3627 (Abell \etal 1989), hereafter the Norma
cluster. No other Abell cluster is located so close to the
Galactic Plane (Andernach 1991).

The overdensity at $(\ell, b) \approx (329\deg, -9\deg)$ is located 
not far from the Norma cluster. It is, however, unrelated to the Norma cluster.
The overdensity is more distant at $v \approx 15\,000$ km s$^{-1}$ (Woudt \etal 1999) and
most likely connected to the X-ray bright Triangulum Australis cluster at
$(\ell, b, v) = (324\deg, -12\deg, 15300$ km s$^{-1}$) (McHardy \etal 1981).

Comparing the extinction contours in Fig.~\ref{cruxgaf2} with the galaxy density contours
in Fig.~\ref{gaf3}, shows that the DIRBE/IRAS reddening map in general is a reliable
tracer of extinction down to very low Galactic latitude. The lowest
galaxy density contour (0.5 galaxies per square degree) compares well with the $A_B = 5^{\rm m}$ extinction
contour. At higher extinction values hardly any galaxies are visible.

\section{The catalogues}

A sample page of the two main catalogues of galaxies in the Crux and GA region is presented in 
Table 1\footnote[1]{Both catalogues in their complete form are available in electronic format at the CDS
via anonymous ftp to cdsarc.u-strasbg.fr (130.79.128.5) or via http://cdsweb.u-strasbg.fr/Abstract.html}.

A short description of the entries in the catalogues is listed below. 

\begin{description}

\item{{\bf Column 1:} WKK running number.}
 
\item{{\bf Column 2:} Second identification.}

\item{{\bf Column 3:} Identification in the IRAS Point Source Catalog (IRAS PSC hereafter). These galaxies are listed 
separately in Table 3. The entries are: I (certain identification), P (possible identification), Q (questionable) and N 
(no credible cross-identification). See Sect.~7 for further discussion.}

\item{{\bf Column 4:} Right Ascension RA (1950.0). }

\item{{\bf Column 5:} Declination Dec (1950.0).}

\item{{\bf Column 6:} Right Ascension RA (2000.0).} 

\item{{\bf Column 7:} Declination Dec (2000.0).}

\item{{\bf Column 8:} Galactic longitude $\ell$.}

\item{{\bf Column 9:} Galactic latitude $b$.}

\item{{\bf Column 10:} Field number of the SRC Survey on which the galaxy was detected. }

\item{{\bf Column 11:} X-coordinate in mm, measured from the centre of the field listed in
column 10. }

\item{{\bf Column 12:} Y-coordinate in mm, measured from the centre of the field listed in
column 10. }

\end{description}

\begin{landscape}
\begin{table}
 \normalsize
 \renewcommand{\baselinestretch}{0.6}
\caption{Galaxies in the southern ZOA -- The Crux Region. Only the first 50 entries are displayed.}
 \label{tab1}
\footnotesize
\begin{tabular*}{24.4cm}{ 
%@{\extracolsep\fill}
  r  @{\extracolsep{1mm}} p{1.8cm} @{\extracolsep{1mm}} 
% 1                       2                      
% RKK                     Other                   
  p{4mm} @{\extracolsep{1mm}}  
% 3
% IR
  l@{\extracolsep{2mm}} l @{\extracolsep{3mm}}
% 4                       5
% RA 1950                 Dec 1950
  l@{\extracolsep{2mm}} l @{\extracolsep{3mm}}
% 4                       5
% RA 2000                 Dec 2000
  r @{\extracolsep{2mm}}r @{\extracolsep{2mm}} r @{\extracolsep{2mm}}
% 6                     7                      8  
% gal l                 gal b                  SRC
  r @{\extracolsep{2mm}}    r @{\extracolsep{2.5mm}} 
% 9                        10                    
% X                        Y                     
  p{6mm} @{\extracolsep{0mm}} p{4.5mm} @{\extracolsep{2mm}}
% 11
% Dx                     d
  r @{\extracolsep{2mm}} r @{\extracolsep{2.5mm}} l @{\extracolsep{1mm}} 
%  12                   13                     14 
%  Bj                   HI                     U
  p{2.7mm} @{\extracolsep{-1mm}} p{2.7mm} @{\extracolsep{-1mm}} 
  p{2.7mm} @{\extracolsep{-1mm}} 
  p{2.7mm} @{\extracolsep{-0.5mm}} p{2.7mm} @{\extracolsep{0mm}} 
% 15a-e      
% T1-5
  p{2.7mm} @{\extracolsep{0.5mm}} p{2.7mm} @{\extracolsep{0mm}} 
% 15f-g      
% T f-g
  l }
%   16
%   Remarks
\hline 
\vspace{-1mm} \\
 WKK & Other & IR & \ \ \ \ R.A. & \ \ \ Dec. & \ \ \ \ R.A. & \ \ \ Dec. & gal $\ell$ \ & gal $b \ $ 
 & SRC & $X$ \ \ & $Y$ \ \ & 
% \multicolumn{2}{c}{D x d}& $B_J}$ & $E(B-V)$ &  & 
 \multicolumn{2}{c}{$D$ x $d$}& $B_{25}$ & $E_{(B-V)}$ &  & 
 \multicolumn{5}{c}{Type} &&& Remarks \\
 Ident & Ident &  & \ \,(1950.0) & \ (1950.0) & \ \,(2000.0) & \ (2000.0) & ($\deg$) \ &($\deg$) \ 
%  & & (mm) & (mm) & \multicolumn{2}{c}{($\arcsec$)} & ($^{\rm m})$ & &
  & & (mm) & (mm) & \multicolumn{2}{c}{($\arcsec$)} & ($^{\rm m})$ & &
  u & \multicolumn{5}{c}{class.} &o &* & \\
\vspace{-1mm} \\
 & & & \hspace{1mm}(h\,\, m\,\, s) & \ ($\deg$\,\, $\arcmin$\,\, $\arcsec$) & \hspace{1mm}(h\,\, m\,\, s) 
& \ ($\deg$\,\, $\arcmin$\,\, $\arcsec$) & & & & & & & & & & & & & & & & & & \\
\vspace{-1mm} \\
 (1) & (2) & (3) & \ \ \ \ (4) & \ \ \ \ (5) & \ \ \ \ (6) & \ \ \ \ (7) & (8) \ & (9) \ 
 & (10) & (11) \ & (12) \ &
 \multicolumn{2}{c}{(13)} & (14) & (15) & & \multicolumn{5}{c}{(16)} 
 &&& (17)\\
\vspace{-1mm} \\
\hline 
\vspace{-1mm} \\
    1 &          &    & 10 08 02.4 & -67 46 31  &  10 09 19.3 & -68 01 17 & 288.64 &  -9.77 &  62 &  -82.8 &  114.2 & \hfill 11x&\hfill  8 & 19.2 & 0.17 & ?&S& & & &?& &1&                                        \\
    2 &          &    & 10 10 31.0 & -68 01 06  &  10 11 48.6 & -68 15 57 & 288.98 &  -9.83 &  62 &  -69.5 &  102.0 & \hfill 13x&\hfill 12 & 18.7 & 0.20 & ?&S& & &L& & &1& vLSB                                   \\
    3 &          &    & 10 12 58.9 & -67 46 15  &  10 14 19.0 & -68 01 11 & 289.03 &  -9.50 &  62 &  -57.8 &  116.0 & \hfill 24x&\hfill  7 & 18.3 & 0.22 & ?&S& & & & &N& &                                        \\
    4 &          &    & 10 16 01.1 & -68 07 12  &  10 17 21.9 & -68 22 14 & 289.47 &  -9.63 &  62 &  -41.7 &   97.9 & \hfill 17x&\hfill  5 & 19.1 & 0.18 &  &S& & & & & & & br.blg                                 \\
    5 & L062-005 &    & 10 16 29.3 & -68 17 24  &  10 17 49.8 & -68 32 27 & 289.60 &  -9.74 &  62 &  -39.1 &   88.9 & \hfill 73x&\hfill 15 & 15.8 & 0.18 &  &S& & &1& &N&S&                                        \\
    6 &          &    & 10 16 42.8 & -67 54 34  &  10 18 04.8 & -68 09 37 & 289.40 &  -9.41 &  62 &  -38.7 &  109.3 & \hfill 32x&\hfill  5 & 18.6 & 0.20 &  &S& & &5&:&E& & vLSBdisk                               \\
    7 &          &    & 10 17 20.8 & -67 52 41  &  10 18 43.3 & -68 07 45 & 289.44 &  -9.36 &  62 &  -35.5 &  111.1 & \hfill 24x&\hfill  7 & 18.0 & 0.20 &  &S& & &M& &E& & p.cov by *                             \\
    8 &          &    & 10 17 52.0 & -68 17 01  &  10 19 13.4 & -68 32 06 & 289.71 &  -9.67 &  62 &  -32.3 &   89.5 & \hfill 30x&\hfill  7 & 17.7 & 0.18 &  &S& & &2& &E& &                                        \\
    9 &          &    & 10 17 53.9 & -68 07 11  &  10 19 15.9 & -68 22 16 & 289.62 &  -9.53 &  62 &  -32.4 &   98.2 & \hfill 23x&\hfill  7 & 18.6 & 0.19 &  &S& & &L& &E&1&                                        \\
   10 &          &    & 10 17 55.5 & -67 55 56  &  10 19 18.2 & -68 11 01 & 289.51 &  -9.37 &  62 &  -32.5 &  108.3 & \hfill 20x&\hfill  5 & 18.8 & 0.20 &  &S& & & & &N& & br.blg or s.p.*                        \\
\vspace{-1.3 mm} \\
   11 &          &    & 10 17 58.1 & -67 59 18  &  10 19 20.6 & -68 14 23 & 289.55 &  -9.41 &  62 &  -32.2 &  105.3 & \hfill 13x&\hfill  8 & 18.5 & 0.20 &  &S& & &0&:& & & br.blg or s.p.*                        \\
   12 &          &    & 10 18 06.7 & -67 57 45  &  10 19 29.4 & -68 12 51 & 289.54 &  -9.39 &  62 &  -31.5 &  106.7 & \hfill 17x&\hfill 12 & 17.9 & 0.20 &  &S& & & &?& & &                                        \\
   13 &          &    & 10 18 07.3 & -68 15 09  &  10 19 29.0 & -68 30 15 & 289.71 &  -9.63 &  62 &  -31.0 &   91.2 & \hfill 16x&\hfill  5 & 18.8 & 0.18 & ?&S& & & & & & & neighb.of 8                            \\
   14 &          &    & 10 18 23.8 & -68 27 04  &  10 19 44.9 & -68 42 10 & 289.84 &  -9.78 &  62 &  -29.4 &   80.6 & \hfill 17x&\hfill 12 & 17.7 & 0.17 &  & & & & & & & & p.cov by **                            \\
   15 &          &    & 10 18 25.5 & -68 19 03  &  10 19 47.1 & -68 34 09 & 289.77 &  -9.67 &  62 &  -29.4 &   87.7 & \hfill 27x&\hfill  8 & 18.5 & 0.17 & ?&S& & & & & & & vvLSB                                  \\
   16 &          &    & 10 19 48.6 & -68 35 55  &  10 21 10.1 & -68 51 04 & 290.03 &  -9.83 &  62 &  -22.3 &   72.8 & \hfill 26x&\hfill  7 & 18.2 & 0.16 &  &S& & & & &E& & br.blg                                 \\
   17 &          &    & 10 20 07.6 & -67 31 25  &  10 21 33.2 & -67 46 34 & 289.46 &  -8.91 &  62 &  -21.9 &  130.5 & \hfill 15x&\hfill  7 & 18.4 & 0.25 & ?&S& & & & & & & cl.to br.*, neighb.of 18               \\
   18 &          &    & 10 20 18.6 & -67 32 40  &  10 21 44.2 & -67 47 50 & 289.49 &  -8.92 &  62 &  -20.9 &  129.3 & \hfill 16x&\hfill  4 & 19.1 & 0.25 &  &S& & & & &E&1& neighb.of 17                           \\
   19 & L062-009 &    & 10 20 36.0 & -68 38 14  &  10 21 57.9 & -68 53 24 & 290.11 &  -9.82 &  62 &  -18.4 &   70.8 & \hfill 38x&\hfill 19 & 16.6 & 0.16 &  &S& & & & & &S& warped?                                \\
   20 &          &    & 10 20 43.1 & -68 08 37  &  10 22 06.9 & -68 23 47 & 289.85 &  -9.40 &  62 &  -18.3 &   97.3 & \hfill 16x&\hfill  8 & 18.6 & 0.19 &  &I& & & &?& &1&                                        \\
\vspace{-1.3 mm} \\
   21 &          &    & 10 20 56.4 & -68 07 13  &  10 22 20.4 & -68 22 24 & 289.86 &  -9.37 &  62 &  -17.2 &   98.5 & \hfill 16x&\hfill 13 & 17.8 & 0.19 &  &S& & &E& & & & br.blg, poss.larger                    \\
   22 &          &    & 10 21 02.6 & -68 03 24  &  10 22 26.9 & -68 18 35 & 289.83 &  -9.31 &  62 &  -16.7 &  102.0 & \hfill 15x&\hfill  8 & 18.4 & 0.21 &  &L& & & & & & &                                        \\
   23 &          &    & 10 21 11.3 & -68 22 50  &  10 22 34.6 & -68 38 01 & 290.02 &  -9.58 &  62 &  -15.7 &   84.6 & \hfill 19x&\hfill  7 & 18.6 & 0.17 &  &S& & & & &E& &                                        \\
   24 &          &    & 10 21 23.2 & -68 32 14  &  10 22 46.0 & -68 47 26 & 290.12 &  -9.70 &  62 &  -14.6 &   76.2 & \hfill 17x&\hfill 11 & 18.4 & 0.16 & ?&S& & & & & & & br.blg or s.p.*                        \\
   25 &          &    & 10 21 26.2 & -67 57 18  &  10 22 51.2 & -68 12 30 & 289.81 &  -9.21 &  62 &  -14.8 &  107.5 & \hfill 27x&\hfill 13 & 17.9 & 0.23 &  &S& & & & & &1& vvLSBdisk                              \\
   26 &          &    & 10 21 42.8 & -68 24 57  &  10 23 06.3 & -68 40 09 & 290.08 &  -9.58 &  62 &  -13.1 &   82.8 & \hfill 26x&\hfill  9 & 18.1 & 0.17 & ?& & & & & & & &                                        \\
   27 &          &    & 10 22 12.1 & -68 12 20  &  10 23 36.7 & -68 27 33 & 290.00 &  -9.38 &  62 &  -10.8 &   94.1 & \hfill 16x&\hfill 15 & 18.2 & 0.18 &  &S& & & & & &1& vLSBdisk                               \\
   28 &          &    & 10 22 20.4 & -68 15 10  &  10 23 44.9 & -68 30 23 & 290.04 &  -9.42 &  62 &  -10.1 &   91.5 & \hfill 17x&\hfill  7 & 18.7 & 0.17 &  &S& & & & & & & br.blg or s.p.*                        \\
   29 &          &    & 10 23 16.4 & -68 29 08  &  10 24 40.7 & -68 44 23 & 290.24 &  -9.57 &  62 &   -5.4 &   79.1 & \hfill 16x&\hfill  7 & 18.8 & 0.17 & ?&S& & & & & & & * s.p. on blg                          \\
   30 &          &    & 10 23 18.1 & -67 59 37  &  10 24 44.2 & -68 14 52 & 289.98 &  -9.15 &  62 &   -5.4 &  105.5 & \hfill 22x&\hfill 13 & 17.8 & 0.22 &  &S& & &L& & &S& br.blg or s.p.*                        \\
\vspace{-1.3 mm} \\
   31 &          &    & 10 23 42.1 & -68 14 47  &  10 25 07.5 & -68 30 03 & 290.14 &  -9.34 &  62 &   -3.3 &   91.9 & \hfill 17x&\hfill 12 & 18.3 & 0.18 &  &S& & & & & & & vLSBdisk                               \\
   32 &          &    & 10 23 43.0 & -68 49 47  &  10 25 06.4 & -69 05 03 & 290.46 &  -9.83 &  62 &   -3.1 &   60.7 & \hfill 15x&\hfill 11 & 18.4 & 0.15 &  &S& & & & & & & br.blg                                 \\
   33 &          &    & 10 23 43.2 & -68 12 49  &  10 25 08.8 & -68 28 05 & 290.13 &  -9.31 &  62 &   -3.3 &   93.7 & \hfill 15x&\hfill  8 & 18.8 & 0.18 &  &S& & &0&:& & & br.blg or s.p.*                        \\
   34 &          &    & 10 23 46.8 & -68 16 00  &  10 25 12.2 & -68 31 16 & 290.16 &  -9.36 &  62 &   -2.9 &   90.9 & \hfill 16x&\hfill 11 & 18.0 & 0.18 &  &S& & & & & &1&                                        \\
   35 &          & N  & 10 24 23.1 & -68 05 26  &  10 25 49.5 & -68 20 43 & 290.12 &  -9.18 &  62 &     .0 &  100.3 & \hfill 13x&\hfill  7 & 18.9 & 0.20 & ?& & & & & & &1&                                        \\
   36 &          &    & 10 26 32.3 & -68 05 04  &  10 28 00.2 & -68 20 25 & 290.29 &  -9.07 &  62 &   10.8 &  100.6 & \hfill 19x&\hfill  7 & 18.2 & 0.20 &  &S& & &0& & & &                                        \\
   37 &          &    & 10 26 56.0 & -67 35 05  &  10 28 25.8 & -67 50 27 & 290.05 &  -8.62 &  62 &   13.0 &  127.4 & \hfill 20x&\hfill 11 & 18.3 & 0.23 &  &S& & &L& & & & vLSBdisk                               \\
   38 &          &    & 10 27 02.6 & -67 46 48  &  10 28 31.9 & -68 02 10 & 290.16 &  -8.78 &  62 &   13.5 &  116.9 & \hfill 23x&\hfill  8 & 18.2 & 0.22 &  &S& & & & & &1&                                        \\
   39 &          &    & 10 27 30.2 & -68 45 01  &  10 28 56.5 & -69 00 23 & 290.71 &  -9.59 &  62 &   15.3 &   64.9 & \hfill 16x&\hfill  8 & 18.5 & 0.18 &  &S& & &M& & & &                                        \\
   40 &          &    & 10 27 59.5 & -67 34 58  &  10 29 30.0 & -67 50 21 & 290.14 &  -8.57 &  62 &   18.4 &  127.4 & \hfill 15x&\hfill  9 & 18.6 & 0.23 & ?& & & & & & & & br.blg or s.p.*                        \\
\vspace{-1.3 mm} \\
   41 &          &    & 10 28 21.3 & -69 03 15  &  10 29 47.1 & -69 18 39 & 290.94 &  -9.81 &  62 &   19.2 &   48.6 & \hfill 20x&\hfill  8 & 18.2 & 0.17 &  &S& & & & &N& & near br.*                              \\
   42 &          &    & 10 28 24.0 & -69 08 50  &  10 29 49.5 & -69 24 14 & 290.99 &  -9.88 &  62 &   19.3 &   43.6 & \hfill 28x&\hfill  5 & 18.0 & 0.17 &  &S& & &0& & &S&                                        \\
   43 &          &    & 10 28 27.0 & -68 55 10  &  10 29 53.4 & -69 10 34 & 290.88 &  -9.69 &  62 &   19.7 &   55.8 & \hfill 19x&\hfill  8 & 18.4 & 0.17 &  &S& & &4& & &1&                                        \\
   44 &          &    & 10 29 17.7 & -69 04 01  &  10 30 44.1 & -69 19 26 & 291.02 &  -9.77 &  62 &   23.7 &   47.8 & \hfill 16x&\hfill 12 & 18.3 & 0.18 & ?&S& & & & & & & near br.*                              \\
   45 &          &    & 10 29 57.9 & -69 13 16  &  10 31 24.3 & -69 28 43 & 291.15 &  -9.88 &  62 &   26.7 &   39.5 & \hfill 19x&\hfill  9 & 17.9 & 0.20 &  &L& & & & & & &                                        \\
   46 &          &    & 10 30 05.7 & -67 29 35  &  10 31 37.9 & -67 45 02 & 290.27 &  -8.39 &  62 &   29.2 &  132.1 & \hfill 20x&\hfill  5 & 18.9 & 0.29 &  &S& & & & &E& & vLSBdisk                               \\
   47 &          &    & 10 30 08.8 & -68 45 22  &  10 31 36.9 & -69 00 49 & 290.93 &  -9.47 &  62 &   28.1 &   64.4 & \hfill 24x&\hfill  5 & 18.5 & 0.19 &  &S& & &4&:&E& &                                        \\
   48 &          &    & 10 32 25.6 & -68 21 18  &  10 33 56.6 & -68 36 49 & 290.90 &  -9.02 &  62 &   39.8 &   85.6 & \hfill 16x&\hfill 11 & 18.6 & 0.19 & ?& & & & & & &1& vvLSB                                  \\
   49 &          &    & 10 32 28.1 & -69 07 24  &  10 33 56.6 & -69 22 55 & 291.30 &  -9.68 &  62 &   38.8 &   44.4 & \hfill 16x&\hfill  5 & 19.4 & 0.20 &  &S& & & & &N& & vLSBdisk                               \\
   50 &          &    & 10 32 38.7 & -69 01 37  &  10 34 07.7 & -69 17 08 & 291.26 &  -9.59 &  62 &   39.8 &   49.6 & \hfill 16x&\hfill  4 & 19.4 & 0.19 &  &S& & & & &E& &                                        \\
\vspace{-1.3 mm} \\
\hline
%\hline
 \end{tabular*}
 \normalsize
\end{table}
\end{landscape}

\noindent
The entries in Table 1 (continued):
\begin{description}

\item{{\bf Column 13:} Large diameter $D$ and small diameter $d$ in arcsec.}

\item{{\bf Column 14:} Apparent magnitude $B_{25}$.}

\item{{\bf Column 15:} The Galactic reddening at the position of the galaxy, as given by the DIRBE/IRAS extinction
maps (Schlegel \etal 1998). See Sect.~5.2 for a more detailed discussion on the calibration
of the Galactic reddening maps at low Galactic latitudes.}

\item{{\bf Column 16:} Morphological type. }

\item{{\bf Column 17:} Remarks.}

\end{description}

\subsection{The Crux region}

In total, 3759\footnote[1]{One galaxy candidate, WKK 1328, is a known Planetary 
Nebula which mistakenly remained in the main catalogue. It is not used in any of the figures and discussion.} 
galaxies were detected with a large diameter $D \ga 0\farcm2$. This list includes 6 galaxies below our diameter
limit. They were included in this list because
of their proximity to (and possibly interacting with)
neighbouring galaxies that do comply with our diameter limit.

\subsubsection{Second identification}

A total number of 88 Lauberts (1982) galaxies were identified in the Crux Region (= 2.3\%).
They are recognisable as 'L' plus the respective field and running number. 
A few of the Lauberts galaxies turned out to be 2 or more individual galaxies after closer
inspection. The Lauberts identification is given in both case, \ie L172--002 = WKK1069 \& WKK1070.
Two galaxies identified by FGCE\# are listed in the Flat Galaxy Catalogue (Karenchentsev \etal 1993),
two in the Arp Madore Catalog (Arp \& Madore 1987) -- recognisable by the AM in Column 2 --,
one in the Parkes-MIT-NRAO 5GHz Radio Survey (Griffith \& Wright 1993, code PMN), and
one in the Catalogue of Southern Ring Galaxies (CSRG) (Buta 1995). A further four galaxies
are listed in the Southern Galaxy Catalogue (Corwin \etal 1985, code SGC) and 11 galaxies
are listed by Visvanathan \& Van den Bergh (1992) from observations of luminous 
spiral galaxies in the direction of the Great Attractor.

\subsubsection{x and y coordinates}

Positive x-values indicate increasing RA, negative x-values 
decreasing RA. Positive y-values point north, negative values 
south with respect to the centre of the field listed in Column 10.
This is not necessarily identical to the field on which that galaxy 
-- based on its coordinates and the actual centre of the survey field --
actually belongs, but on which it was first identified in the course of this survey.

\subsubsection{Morphological types}

The morphological types are coded similarly to the 
precepts of the Second Reference Catalogue (de Vaucouleurs \etal 1976). However, due to 
the varying foreground extinction, a homogeneous and detailed type classification could 
not always be accomplished and some codes were added (see Paper I for details).

\subsubsection{Descriptive remarks}

The remarks and the abbreviations in column 17 are generally selfexplanatory.
The most common abbreviations are already explained in Paper I. There are, however, some 
additional abbreviations used in this paper:

\begin{tabular}{lp{6cm}}
centr. & central \\
dw. & dwarf \\
ext. & extended \\
pec. & peculiar \\
perp. & perpendicular \\
pl.flaw & plate flaw\\
sm. & small\\
sh.edge & sharp edge \\
\end{tabular}

\subsection{The Great Attractor region}

In the GA region, 4423 galaxies (WKK 3761 -- WKK 8183) were detected with their major axis $D \ga 0\farcm2$ (4
of which have $D < 0\farcm2$). The catalogue of the GA region has the exact same format as the Crux catalogue.

\subsubsection{Second identification}

Most of the second identifications originate from the ESO/Uppsala Survey of the
ESO(B) Atlas (Lauberts 1982), recognisable as `L' plus the respective field and running
number. A total number of 108 Lauberts galaxies were identified in the Great Attractor 
region (= 2.4\%). After closer inspection a number of the Lauberts galaxies turned out
to be 2 or more individual galaxies. The Lauberts identification is given in both cases,
\ie L138-012 = WKK7813 and WKK7815. Two galaxies indicated by FGCE\# are listed in
the Flat Galaxy Catalogue (Karenchentsev \etal 1993), three in the Catalogue of Southern
Ring Galaxies (CSRG) (Buta 1995), and two radio galaxies were previously recorded (Jones
and McAdam 1992, code PKS and JM). A further 8 galaxies are listed in the Southern
Galaxy Catalogue (Corwin \etal 1985, code SGC).

\section{On the calibration and reliability of the estimated $B_{25}$ magnitudes and diameters}

For the calibration of the estimated magnitudes, the same procedure was followed as presented
in Paper I. For each of the 37 eye-balled film copies, the recorded surface brightnesses were 
calibrated using the $\log(I)$ intensity scale on the film copies.

Care was taken to scan small overlapping areas on each of the surveyed film copies
to check for variations in the zero point of the magnitude estimate from field to field. 
This comparison revealed that the magnitudes of galaxies in 
three fields in the Crux region were systematically offset, namely for F65, F98 and
F132. Their zero point differences are given in Table~\ref{zpo}. A correction was made for
these fields, so that all the magnitudes in our catalogue are internally consistent.

%\addtocounter{table}{+1}
\begin{table}[h]
 \caption{Field to field variations in the zero points of the magnitude estimates.}
 \label{zpo}
\begin{center}
\begin{tabular}{ccclc}
\hline
\vspace{-2.3mm} \\
Field &  $<B_{25}>$ & $<D>$  & Remarks \hspace*{0.25cm} & N \\
\vspace{-1.3mm} \\
\hline
\vspace{-1.3mm} \\
 065  & \hfill  --0$\fm$35 $\pm$ 0$\fm$44 & \hfill   4$\arcsec   \pm  6\arcsec$  & Too bright & 20 \\
 098  & \hfill   +0$\fm$35 $\pm$ 0$\fm$44 & \hfill  No deviation                 & Too faint & 15 \\
 132  & \hfill   +0$\fm$45 $\pm$ 0$\fm$37 & \hfill $-4\arcsec    \pm  6\arcsec$  & Too faint & 12 \\
\vspace{-1.3mm} \\
\hline
\end{tabular}
\end{center}
\end{table}

\subsection{Magnitude calibration}

\subsubsection{The Crux region}

Galaxies in common with the ESO--LV catalogue (Lau-berts \& Valentijn 1989) have been used to 
calibrate our magnitude estimates. Unfortunately only 15 ESO--LV ga-laxies are present in the
Crux region: 12 galaxies on F171, one galaxy on F173, one galaxy on F175 and one galaxy on F97 
(the large, nearby Circinus galaxy). 

\begin{figure}
 \resizebox{\hsize}{!}{\includegraphics{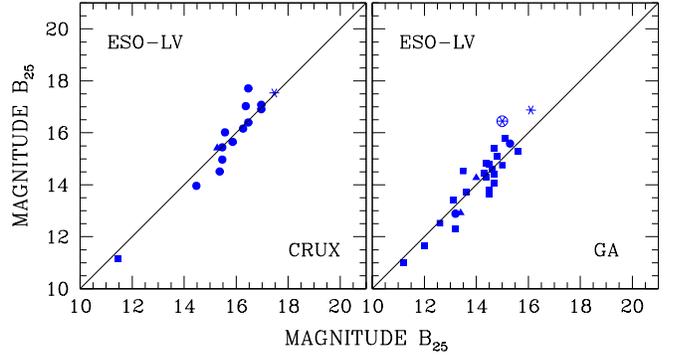}}
\caption{The comparison of the observed magnitudes with the ESO--LV catalogue
in the Crux region (left panel) and the Great Attractor region (right panel)
Note that the line drawn in each of the
diagrams is not a fit to the data.}
\label{magcal}
\end{figure}

In Fig.~\ref{magcal} we show the magnitude comparison for galaxies which have a counterpart
in the ESO--LV catalogue. Our magnitude estimate compares best with the $B_{25}$ isophotal
magnitude of the ESO--LV magnitude (see top-left panel of Fig.~\ref{magcal}). In our catalogues,
and our papers, our magnitude estimates refer to $B_{25}$ of the ESO--LV catalogue (Lauberts \& Valentijn 1989),
and not to Johnson $B_J$. There is a reported $\sim$0.2 mag difference between these two photometric
systems (Peletier \etal 1994; Alonso \etal 1993).

The galaxies with a counterpart in the ESO--LV catalogue are indicated by different 
symbols in the left panel of Fig.~\ref{magcal}; circles, stars, triangles and squares 
for F171, F173, F175 and F97 respectively.
The overall agreement in the derived magnitudes is good with no systematic deviations from linearity 
over the entire range of magnitudes (including even the fainter galaxies).
The uncertainty in the derived magnitude estimate (from the ESO--LV comparison) is very good for
eye-estimated magnitudes, namely 1$\sigma = 0\fm51$.
Note that our magnitude estimate is an isophotal magnitude, not the total magnitude of a galaxy.

\subsubsection{The Great Attractor region}

For the magnitude calibration in the Great Attractor region, the same
procedure was followed. In the GA region, 27 galaxies are listed in the ESO--LV catalogue: 2 galaxies
on F136, 21 galaxies on F137, 2 galaxies on F138 and 2 galaxies on F221. 
In the right panel of Fig.~\ref{magcal} we show the results of this calibration; our
calibrated magnitude is plotted against the $B_{25}$ of the ESO--LV catalogue. The filled circles,
squares, triangles and stars correspond to galaxies on fields F136, F137, F138 and F221, respectively.
Note, that one of the galaxies on F221, indicated by the encircled star symbol in the right 
panel of Fig.~\ref{magcal}, has not been included in the calculation of the mean difference
in magnitude and diameter. The $B_{25}$ entry in the ESO--LV catalogue is questionable for this
object.

The correspondence in magnitudes is good with no deviations from linearity 
over the {\sl entire} magnitude range. The scatter in the magnitude data in the GA region, $ 1 \sigma = 0\fm51$, is identical 
to the observed scatter in the Crux region. 

\subsection{Diameter comparison}

\begin{figure}
 \resizebox{\hsize}{!}{\includegraphics{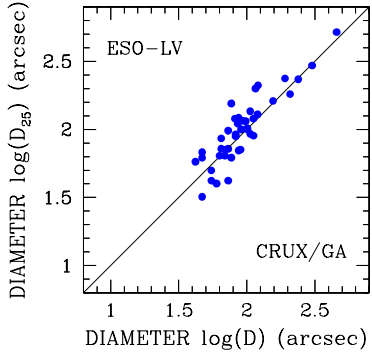}}
\caption{The comparison of the observed diameters with the ESO--LV catalogue ($D_{25}$)
in the Crux and Great Attractor regions. Note that the line drawn in the
diagram is not a fit to the data.}
\label{diacom}
\end{figure}

The diameter comparison (with $D_{25}$ of ESO--LV) reveals a somewhat larger scatter 
(see Fig.~\ref{diacom}) and suggests that our diameter estimates corresponds 
to a higher isophotal level, i.e.~at $B$=24.5 mag arcsec$^{-2}$. This is entirely consistent with 
the findings in the Hydra/Antlia region (Paper I). From this comparison -- combining the data in the
Crux and GA regions -- the following statistics are
derived:

$$D ({\rm Crux/GA}) - D_{25} ({\rm ESO\,LV}) = -11{\arcsec} \pm 29{\arcsec} $$

There is no significant difference in these statistics if the Crux
and GA regions are examined individually, namely

$$D ({\rm Crux}) - D_{25} ({\rm ESO\,LV}) = -9{\arcsec} \pm 22{\arcsec}$$
$$D ({\rm GA}) - D_{25} ({\rm ESO\,LV}) = -14{\arcsec} \pm 31{\arcsec} $$

\subsection{Internal comparison}

\begin{figure}
 \resizebox{\hsize}{!}{\includegraphics{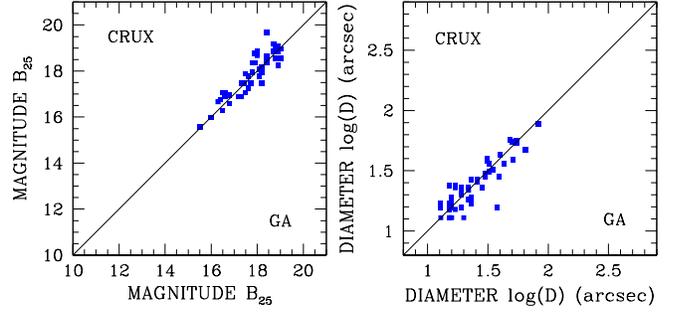}}
\caption{An internal comparison between the observed magnitudes (left) and diameters (right)
in the Crux and Great Attractor regions. The line drawn in each of the
diagrams is not a fit to the data.}
\label{cruxgafig6}
\end{figure}

The overlap of galaxies between the Crux region and the GA region allows a check on the internal consistency.
There are 49 galaxies on the border of the Crux and GA region: 4 galaxies on F67/F99, 6 galaxies on F98/F99, 
23 galaxies on F174/F221, 6 galaxies on F175/F176, 2 galaxies on F175/F221 and 8 galaxies on F175/F222. 
These data are displayed in Fig.~\ref{cruxgafig6}. The data clearly show that the magnitudes
and diameters in the Crux and GA region are fully consistent. The scatter in the observed magnitude is 
low (0$\fm$41). A small offset in the magnitudes ($-0\fm$12) is visible. The diameters show no offset and 
the observed scatter is very low (6$\arcsec$).
 
$${B_{25}} ({\rm GA}) - {B_{25}} ({\rm Crux}) = -0{\fm}12 \pm 0{\fm}41 $$
$${D} ({\rm GA}) - {D} ({\rm Crux}) = 0{\arcsec} \pm 6{\arcsec} $$

\subsection{Other Zone of Avoidance catalogues}

\begin{figure}
 \resizebox{\hsize}{!}{\includegraphics{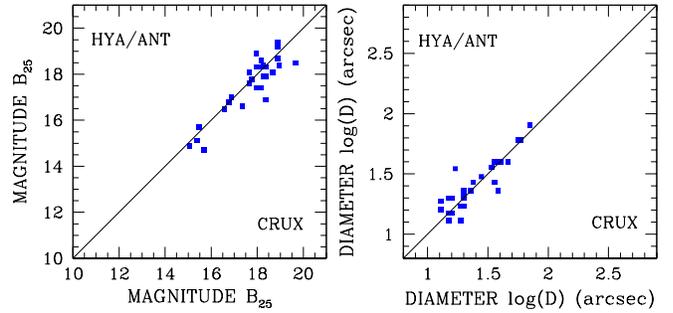}}
\caption{The comparison of the observed magnitudes (left) and diameters (right)
in the Crux region with the Hydra/Antlia region (Paper I).
The line drawn in each of the diagrams is not a fit to the data.}
\label{gafig4}
\end{figure}

A comparison was made for galaxies found in both the Crux region and the 
adjacent Hydra/Antlia region (Paper I). This is the only other ZOA galaxy catalogue
at the moment for which a quantitative comparison is possible. There are 28 galaxies in common, spread over the 
borders of four fields; 10 galaxies on F171/F170, 14 galaxies on F62/F92, 2 galaxies on
F62/F93 and 2 galaxies on F63/F93. These data are shown in Fig.~\ref{gafig4}.
Again the agreement in magnitude is good with a similar scatter around the mean
as for the ESO--LV galaxies, and with no 
systematic deviations from linearity over the displayed magnitude range. A slight offset in 
the magnitude zero point is evident, the galaxies presented in Paper I in the Hydra/Antlia 
region are on average 0$\fm$17 brighter. The scatter (0$\fm$55) is identical to the scatter in 
the ESO--LV data (0$\fm$51). The comparison of the major diameters in the Crux and Hydra/Antlia 
region, reveals a much lower scatter (6$\arcsec$) compared to the equivalent ESO--LV analysis (22$\arcsec$). 
The tighter relationship implies that the Crux estimates are indeed measured to the same isophotal 
level $B$=24.5 mag arcsec$^{-2}$ as the earlier survey (Paper I).

$$B_{25} ({\rm Crux}) - B_{25} ({\rm Hya/Ant}) = 0{\fm}17 \pm 0{\fm}55 $$
$$D ({\rm Crux}) - D ({\rm Hya/Ant}) = -1{\arcsec} \pm 6{\arcsec} $$

Overall, these statistics show that there are no systematic differences between the Crux and the 
Hydra/Antlia catalogues, despite the fact that the scanning of the sky survey plates were performed 
by two different persons (PAW for the Crux and GA region and RCKK for the Hydra/Antlia region).
Together, these surveys therefore form a uniformly defined catalogue of partially obscured galaxies 
in the southern Milky Way.

The two catalogues of partially obscured galaxies in the southern Milky Way, \ie the Hydra/Antlia region (Paper I) and the 
Crux and GA regions are internally consistent and can therefore be merged to form a large, uniformly defined
catalogue of galaxies in the southern ZOA, spanning $\sim75\deg$ in Galactic longitude and $\sim20\deg$ in Galactic 
latitude. The observational evidence for this statement are listed below:
 
\begin{itemize}
\item[$\bullet$]{The identical scatter in the comparison with the ESO--LV magnitude: 0$\fm$46 (Hydra/Antlia, Paper I), 
0$\fm$51 (Crux) and 0$\fm$51 (GA).}
\item[$\bullet$]{The small offset in the magnitude comparison with other ZOA catalogues: 0$\fm$17 (Crux vs.~Hydra/Antlia) and $-0\fm12$ (GA vs.~Crux).}
\item[$\bullet$]{The negligable offset and consistent scatter in the major diameter estimates: $1 \sigma = 6\arcsec$ for both the 
Crux vs.~Hydra/Antlia and GA vs.~Crux comparison.}
\end{itemize}

\section{The Galactic foreground extinction}

A detailed analysis of the Galactic foreground extinction is presented by Woudt (1998). 
Here, the two most common extinction indicators are briefly discussed.

\subsection{The neutral hydrogen in the Milky Way} 

Assuming no variation in the gas-to-dust ratio, the neutral hydrogen (HI) content in the Milky Way
can be used as an indicator of the foreground extinction.
Close to the plane of the Milky Way (within $1{\deg}-2\deg$ of the Galactic Equator),
the Galactic HI line might be saturated, thereby
underestimating the true extinction. At these latitudes the Galactic CO (Dame \etal 1987) might be a 
better tracer. However, as we have found no galaxy candidates that close to the Galactic Plane, the HI
column densities would in principle be adequate.

Following the precepts of Burstein \& Heiles (1978, 1982), 
the Galactic foreground extinction $A_B$ = $1.337 \times A_V =
4.14 \times E_{(B-V)}$ (Cardelli \etal 1989) can be determined from the 
Galactic HI column density alone by

\begin{equation}    
 E_{(B-V)} = \left({{N({\rm HI})}\over{2.23 \cdot 10^{18}}}\right) \times 4.43 \cdot 10^{-4} - 0.055 \\
\end{equation}

\noindent
where $N({\rm HI})$ is in units of 10$^{21}$ atoms cm$^{-2}$.
However, the gas-to-dust ratio {\sl does} vary locally. As discussed in Woudt (1998),
we support the earlier finding by Burstein \etal (1987), that the HI gas-to-dust in the Crux
and Great Attractor region is twice the
nominal value. A higher gas-to-dust ratio locally implies an overestimation of
the Galactic extinction.

\subsection{The 100 micron DIRBE/IRAS extinction maps} 

Schlegel \etal (1998)
have presented the 100 micron extinction maps from the DIRBE
experiment. These maps have a much better angular resolution
(6$\farcm$1) compared to the HI maps ($\sim$20--30 arcminutes) and
variations in the gas-to-dust ratio are not important. These maps
provide a direct measure of the dust column density.
The reddening maps, however, have only been calibrated by the colours of 
elliptical galaxies ($(B-R)$ and $(B-V)$, see Schlegel \etal (1998) for details) 
at intermediate and high Galactic latitude.

Woudt (1998) has tested the calibration
of the reddening maps at low latitude using the ($B-R$) colour and Mg$_2$ index 
of 18 elliptical galaxies in the Crux and Great Attractor region. The agreement
between the reddening values of the DIRBE/IRAS extinction maps and the $E_{(B-V)}$
values derived from the colour-Mg$_2$ relation (Bender \etal 1993) is generally good (Woudt 1998).
Note, however, that because of the choice of $B$ and $R$ filters the latter values are very sensitive 
to the assumed Galactic extinction law ($R_V = 3.1$). Large systematic uncertainties could 
therefore arise when deviations from the mean Galactic extinction law occur. 
A different choice of filters, e.g. $R$ and $K'$, eliminates this uncertainty because the Galactic
extinction law is insensitive to the assumed value of $R_V$ in the near infrared (see Cardelli
\etal 1989).
A calibration of the reddening maps at low Galactic latitude using the ($R-K'$) colour of low-latitude
elliptical galaxies is highly desirable. 
For more details on the problem of selecting the `right' $R_V$ value, see
McCall \& Armour (2000).

In this paper, we exclusively use the DIRBE/IRAS reddening maps for 
extinction corrections. Furthermore, we assume $R_V = 3.1$ for the conversion of selective
extinction to total extinction, and $A_B/A_V = 1.337$ (Cardelli \etal 1989).

\section{Properties of the obscured galaxies}
 
\subsection{The Crux region}

\begin{figure}
 \resizebox{\hsize}{!}{\includegraphics{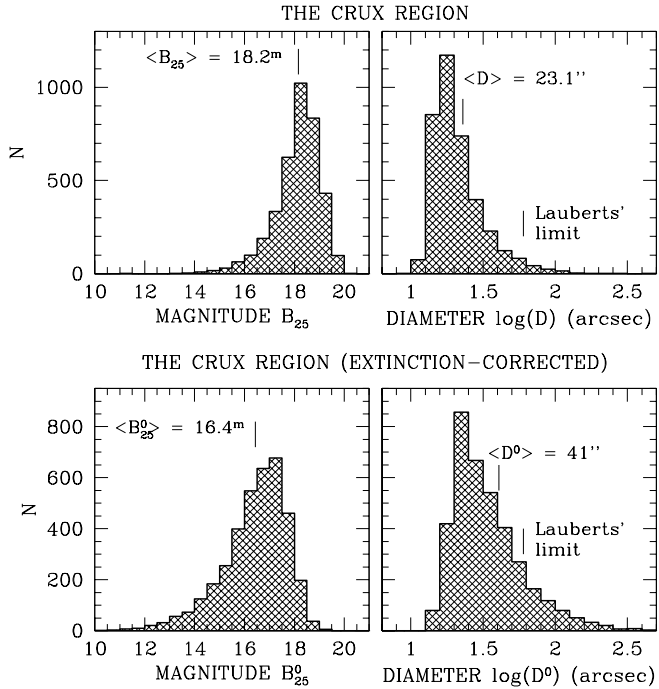}}
\caption{The distribution of the magnitudes (left panel) and diameters (right panel) of 
the 3759 galaxies discovered in the Crux region, with observed values in the top panel 
and extinction-corrected values in the bottom panel.}
\label{cruxbjd}
\end{figure}

The top panels of Fig.~\ref{cruxbjd} show the distribution of the observed magnitudes (left) and diameters 
(right) of the 3759 galaxies in the Crux region. On average the galaxies are quite small 
($<D>$ = 23$\farcs$1) and faint ($<B_{25}>$ = 18$\fm$2); nearly  
identical to what was found in the Hydra/Antlia
region (Paper I), \ie 21$\farcs$8 and 18$\fm$2.
The above graph indicates that our survey is fairly complete for galaxies 
greater than $D \ga 24\arcsec$ or brighter than $B_{25} \la 18^{\rm m}$.
As the diameters and magnitudes in this diagram are heavily influenced by the obscuring 
effects of the Milky Way, these numbers should be regarded as indicative only.

In the Crux region, 106 galaxies have a recorded major diameter $D \ge 60\arcsec$ 
(the Lauberts (1982) diameter limit which is indicated in Fig.~\ref{cruxbjd}). Of the 88 galaxies recorded 
by Lauberts in the Crux region, 22 (= 25\%) are actually smaller than 60$\arcsec$. 
In almost all cases, these galaxies are either covered by many stars, or are part of a double/triple system.
Nonetheless, 40 galaxies (40/106 = 38\%) with $D \ge 60\arcsec$ were not previously recorded by Lauberts. These statistics 
improve somewhat in favour of the Lauberts catalogue for galaxies larger than $1\farcm3$, the diameter 
limit for which the ESO--LV catalogue is said to be complete (Hudson \& Lynden-Bell 1991). 
Still, 10 galaxies out of the 45 galaxies larger than $1\farcm3$ (= 22\%) were not identified by Lauberts.

Using the DIRBE/IRAS reddening values (Column 15 in Table 1) the observed parameters
(diameters and magnitudes) can be corrected for absorption following Cameron (1990).
We then obtain the distribution of extinction-corrected magnitudes ($B^0_{25}$) and 
diameters ($D^0$) as shown in the lower panels of Fig.~\ref{cruxbjd}.
Extinction-corrected diameters of the galaxies in the deepest extinction layers seem 
unrealistically large. The Circinus galaxy, WKK3050, with $A_B = 5\fm72$ has an observed
diameter of 457$\arcsec$ ($\log D = 2.66$). After the extinction correction, the diameter of the 
Circinus galaxy would be 11.1 degrees ($\log D^0 = 4.60$)! This galaxy is not displayed in Fig.~\ref{cruxbjd}. 
Because of this, a more detailed study of the obscurational effects on the magnitudes
and diameters of ZOA galaxies is desired, especially for high extinction ($A_B > 3^{\rm m}$).

An additional complication arises for the Circinus galaxy, as well as for other strong IRAS galaxies.
This nearby Seyfert 2 galaxy is a strong IRAS source ($f_{100} = 315.90$ Jy).
This most definitely influences the derived Galactic reddening value by Schlegel \etal (1998) at the location of Circinus,
as only 5 arcminutes from the centre of Circinus, the extinction drops to $A_B = 3\fm56$, a full $2^m$ lower. 
One should be aware when using the Schlegel \etal (1998) reddening maps
that local galaxies -- depending on the flux -- can in fact contaminate the reddening values. Care should be taken
when using these extinction values at face-value.

The average, extinction-corrected magnitudes and diameters for the galaxies in the 
Crux region are $B^0_{25} = 16\fm4$ and $D^0 = 41\arcsec$. These means have been determined
for the magnitude and diameter range displayed in the lower panels of Fig.~\ref{cruxbjd}, not including those galaxies 
that might have been overcorrected, \ie only galaxies with an extinction correction of $\Delta m \le 6^{\rm m}$ are
used. Note also that the extinction correction does not take the patchiness in the distribution 
of, for instance, local dust clouds into account. 

A total of 593 galaxies in the Crux region have extinction-corrected diameters larger or equal than
$60\arcsec$, \ie the Lauberts (1982) diameter limit. This means that in the absence of the 
obscuration by the Milky Way, Lauberts would have detected 593 galaxies instead of 
the recorded 88 galaxies. The diminishing effects of the Galactic foreground extinction
could not have been more clearly illustrated.

\subsection{The Great Attractor region}

\begin{figure}
 \resizebox{\hsize}{!}{\includegraphics{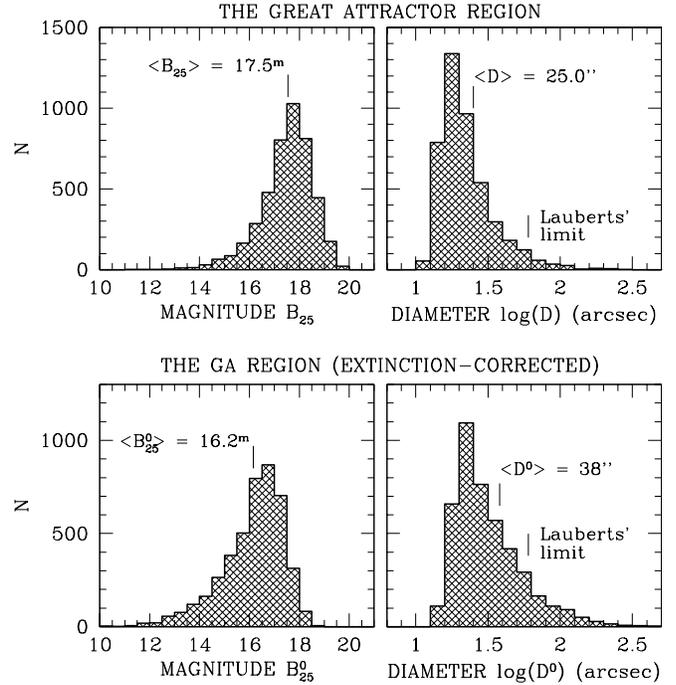}}
\caption{The distribution of the observed magnitudes (left panel) and diameters (right panel) of 
the 4423 galaxies discovered in the Great Attractor region.}
\label{gafig5}
\end{figure}

The upper panels of Fig.~\ref{gafig5}  display the distribution of the observed apparent 
magnitude ($B_{25}$) and major diameter ($D$) of the galaxies unveiled in the GA region. The observed 
characteristics of the galaxies in the GA region are significantly different from both the Crux and 
Hydra/Antlia region, see also Sect.~6.4. On average, the galaxies are somewhat larger ($<D> = 25\farcs0$)
and much brighter ($<B_{25}> = 17\fm5$) compared to the Crux region ($23\farcs1$ and $18\fm2$) and the
Hydra/Antlia region ($21\farcs8$ and $18\fm2$).  
This either means that the Galactic foreground extinction is much lower in the GA region (to explain
for the fact that the galaxies are on average 0.7 mag brighter in the GA region), or,
alternatively, the GA region encompasses a nearby overdensity of galaxies. 

From the Galactic reddening maps (Schlegel \etal 1998) there is some evidence that the mean extinction
for galaxies in the Crux region is somewhat higher ($0\fm2$) compared to galaxies in the GA and Hydra/Antlia regions 
(see the large excursion of the $A_B = 1^{\rm m}$ contour below the Galactic Plane
at $295\deg \la \ell \la 305\deg$ in Fig.~\ref{cruxf1}).
This difference is not enough, however, to fully explain
why galaxies in the GA region are 0.7 mag brighter, especially since galaxies in the Hydra/Antlia
region are subjected to approximately the same extinction as galaxies in the GA region. 
Therefore both effects seem to have influenced the GA sample (see Sect.~6.4).

In the GA region, 161 galaxies have a major diameter $D \ge 60\arcsec$, of which 95 had been recorded
previously by Lauberts (1982). 13 of the galaxies found by Lauberts (1982) have diameters smaller than
60$\arcsec$. In the Crux region, 38\% of the galaxies larger
than $60\arcsec$ had not been recorded by Lauberts, in the GA region this incompleteness is 40\%.

The ``extinction-corrected'' magnitude and diameter distributions of the galaxies in the GA region
are shown in the lower panels of Fig.~\ref{gafig5}. 
The mean, extinction-corrected magnitude and diameter of the galaxies in the GA region
is $<B_{25}^0> = 16\fm2$, and $D^0 = 38\arcsec$, respectively. These values have been determined based on the
galaxies displayed in the lower panels of Fig.~\ref{gafig5}, \ie not including the galaxies in the
deepest layers of the foreground extinction for which the correction becomes increasingly uncertain. 
Galaxies with a total extinction correction of $\Delta m \ge 6$ mag (see Cameron, 1990) 
are not included in these statistics.

A total of 584 galaxies have extinction-corrected diameters larger or equal than 60$\arcsec$, \ie the
Lauberts (1982) diameter limit. In the absence of the obscuration by the Milky Way, Lauberts would have detected 
584 galaxies within the limits of our survey, instead of the recorded 108 galaxies. Again, these numbers demonstrate 
the incompleteness (= 82\%) in the Lauberts catalogue near the plane of the Milky Way.

\subsection{The completeness of our survey}

\begin{figure}
 \resizebox{\hsize}{!}{\includegraphics{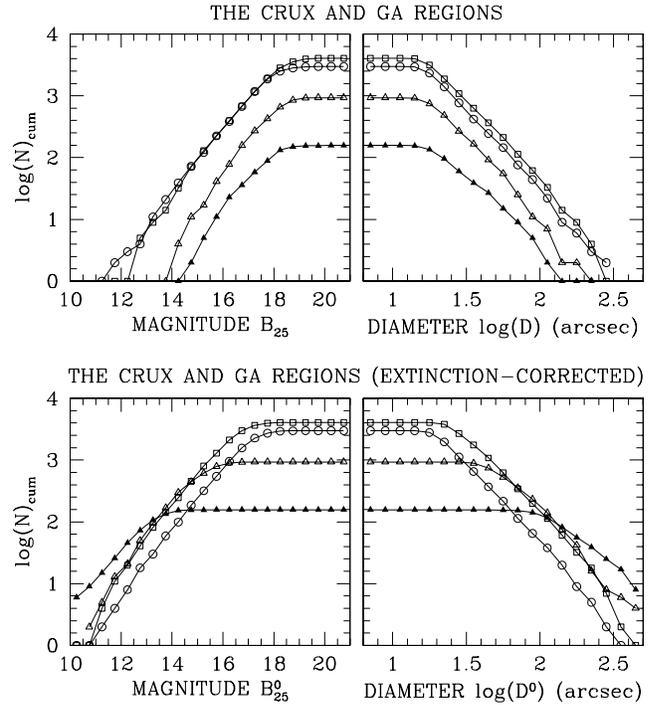}}
\caption{The cumulative distribution of observed (top panels) and extinction-corrected
(lower panels) magnitudes (left) and diameters (right) for four different intervals of Galactic
foreground extinction. The open circles display the galaxies with $A_B \le 1^{\rm m}$, the open
squares are galaxies with $1^{\rm m} < A_B \le 2^{\rm m}$, the open triangles correspond to galaxies
with $2^{\rm m} < A_B \le 3^{\rm m}$, and the filled triangles are galaxies with $3^{\rm m} < A_B \le 4^{\rm m}$. }
\label{cruxcompl}
\end{figure}

The cumulative magnitude and diameter curves, $B_{25} - \log N_{cum}$ and $\log D - \log N_{cum}$, 
plotted in Fig.~\ref{cruxcompl} allow us to assess the completeness of our optical survey. 
For this analysis, we have taken all the galaxies in the Crux region and the Great Attractor region
together.
The four different curves in Fig.~\ref{cruxcompl} illustrate four intervals 
in Galactic foreground extinction. In the interval
$0\fm45 \le A_B \le 1^{\rm m}$ (open circles in Fig.~\ref{cruxcompl}), 2978 galaxies are present. 
4045 galaxies are located in the next interval ($1^{\rm m} < A_B \le 2^{\rm m}$, open squares), 
932 galaxies in $2^{\rm m} < A_B \le 3^{\rm m}$ (open triangles), and 157 galaxies are present in
the interval $3^{\rm m} < A_B \le 4^{\rm m}$ (filled triangles).

In the top panels which shows the ``observed'' diameter and magnitude distribution, 
one can see a linear increase in the cumulative number distribution for $A_B \le 3^{\rm m}$ up 
to $B_{25} = 18\fm0 - 18\fm5$ and $\log D \approx 1.2$ ($D \approx 16''$), 
after which the curves begin to flatten. These numbers give a fair indication of the
completeness limits of the observed parameters of our survey and compare well with the
completeness limits found in the Hydra/Antlia region ($B_{25} = 18\fm5$ and $D = 14''$
for $A_B \le 3^{\rm m}$, see Paper I).

\begin{figure}
 \resizebox{\hsize}{!}{\includegraphics{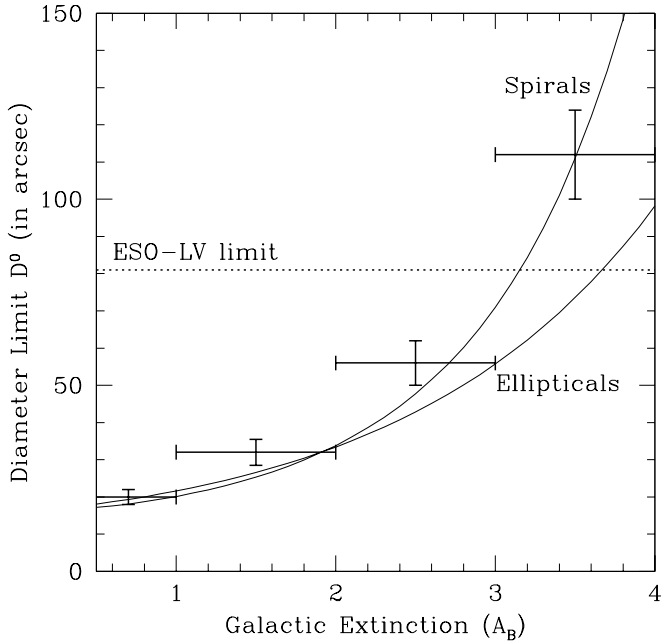}}
\caption{The extinction-corrected diameter completeness limit as a function of the Galactic
foreground extinction. The horizontal bars indicate the completeness limit over the range this limit
was determined. The vertical bars show the uncertainty in the completeness limit.}
\label{dialim}
\end{figure}

In terms of merging our optical catalogue with existing catalogues, we should, however, consider
the completeness limits of the ``extinction-corrected'' parameters (see the bottom panels of Fig.~\ref{cruxcompl},
where one can read of the completeness limits once the curves begin to flatten).
The two curves in Fig.~\ref{dialim} show how our {\sl observed} diameter completeness limit
of $D = 16''$ translates into an extinction-corrected diameter completeness limit for
elliptical and spiral galaxies, respectively, as a function of the Galactic foreground extinction,
following the precepts of Cameron (1990).
The horizontal bars in Fig.~\ref{dialim} correspond to the completeness limits derived from the four different 
intervals displayed in the lower-right panel of Fig.~\ref{cruxcompl}, indicating the range of which they were 
determined. The vertical bars show the error in the determined completeness limit.

The curves illustrate that at higher extinction, the diameter correction increases dramatically. An uncertainty
of 0.1 mag in the Galactic foreground extinction translates into an uncertainty of 7\%, 8\% and 11\%
in the diameter correction for $A_B$ = 2$^{\rm m}$, 3$^{\rm m}$ and 4$^{\rm m}$, respectively. For the extinction-correction
we use the galaxy classification given in Tables 1 and 2. As the dominant fraction of
the galaxies found in our survey are spirals -- the mixture of galaxy types in the
Crux and GA region is (E-S0 : S-I : unclassified) $\approx$ (10\% : 75\% : 15\%) --
it is to be expected that our measured diameter completeness limit follows the `spiral' curve
closely.

Despite the uncertainties in the extinction correction, Fig.~\ref{dialim} shows that we are complete for all 
galaxies (spiral and elliptical) with $D^0 \ge 32''$ for $A_B \le 2^{\rm m}$
and that we are complete for all galaxies (ellipticals and spirals) with $D^0 \ge 1\farcm3$ for $A_B \le 3^{\rm m}$, the
ESO-LV diameter completeness limit. This, again, is consistent with the results in
the Hydra/Antlia region (Paper I). 

\subsection{An overdensity in the Great Attractor region}

\begin{figure}
 \resizebox{\hsize}{!}{\includegraphics{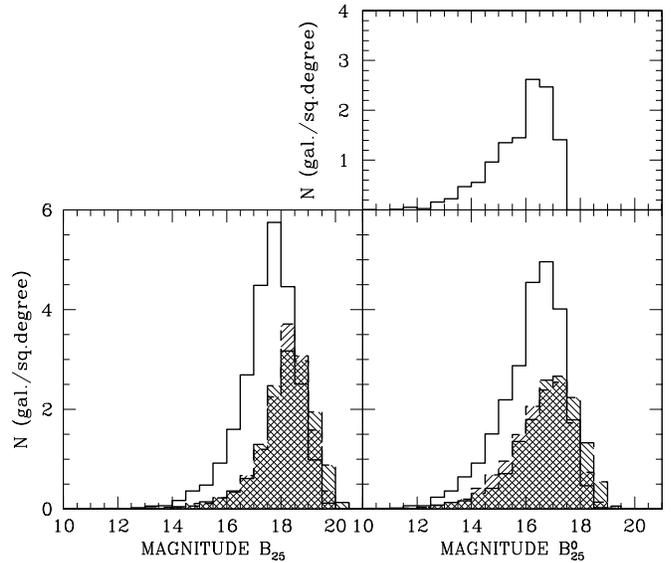}}
\caption{A comparison between the observed magnitude distribution (normalised by the area covered)
in the Hydra/Antlia, Crux and Great Attractor regions for galaxies with $A_B \le 3^{\rm m}$. The Hydra/Antlia region is
indicated by the backward slanting histogram, the Crux region by the forward slanting histogram and the
Great Attractor region by the open histogram. The top-right histogram shows the difference between the galaxy density
in magnitude bins of 0.5 mag in the Great Attractor region, and the mean of galaxy density the Crux and Hydra/Antlia region.}
\label{gafig5b}
\end{figure}

In Fig.~\ref{gafig5b}, we compare the magnitude distribution in the Great Attractor region with that
of the Hydra/Antlia and Crux region for all the galaxies with $A_B \le 3^{\rm m}$. The distribution is
given in number of galaxies per square degree in bins of 0.5 mag. At extinction levels $A_B \le 3^{\rm m}$,
the total area covered in the Hydra/Antlia, Crux and Great Attractor region is approximately
223, 266 and 175 square degrees, respectively. Within these limits, there are 3227, 3629 and 4326 galaxies present
in our catalogues.

The distribution of the extinction-corrected magnitudes (lower-right panel of Fig.~\ref{gafig5b}
of the Hydra/Antlia and Crux galaxies match surprisingly well, despite the different large-scale
structures that have been observed in these regions (Kraan-Korteweg \etal 1995; Fairall \etal 1998).
The extinction-corrected magnitude distribution in the Great Attractor region is noticeably 
different, it reveals a significant {\sl excess of galaxies in the Great Attractor region}. 

In the histogram in the top panel of Fig.~\ref{gafig5b} the difference of the GA distribution 
with the mean of the Crux and Hydra/Antlia profiles (lower-right panel) is shown. 
The distribution (shape) of the excess galaxies is similar to the magnitude 
distribution of galaxies in the Coma cluster (data from Godwin et al., 1983) 
for $B^0_{25} \le 16\fm0$, except for an offset of the histogram of
$\sim$1 magnitude. Hence, this excess could (in part) be explained by a rich cluster of
galaxies present in the Great Attractor region, \ie the Norma cluster. 
The galaxies in Coma are (on average) one magnitude fainter than the 
excess galaxies in the Great Attractor region which puts the excess
galaxies in the distance range of the Great Attractor, i.e., $\sim$1.6 times closer than Coma, 
implying a redshift-distance of $\sim$4250 km s$^{-1}$. Given the uncertainties in the extinction correction,
this number should be regarded as tentative only.

Despite the uncertainties -- most notably the extinction correction -- the magnitude
distribution in the GA region alone provides strong evidence for a significant 
{\sl excess of galaxies} belonging to the Great Attractor, a large fraction of which is due to the
Norma cluster.

It should be noted that the galaxies in the vicinity of the Norma cluster are of distinctly different morphological mix as
compared to the overall mixture of galaxy types. Overall, the galaxy mixture is (E/S0 -- Spiral/Irregular -- Uncertain) $\approx$ 
(10\% -- 75\% -- 15\%), but
within the Abell radius of the Norma cluster ($1.75\deg$ at the distance of the Norma cluster; Woudt, 1998) the galaxian mix is 
(E/S0 -- S/I -- Uncertain) = (20\% -- 68\% -- 12\%), and within the core radius of the Norma cluster ($10\farcm4$; Woudt, 1998)
the mixture of galaxy types is (E/S0 -- S/I -- Uncertain) = (48\% -- 48\% -- 4\%).
These numbers indicate that the Norma cluster is a rich cluster, dominated by a large population of early-type galaxies.

\section{Galaxies in the IRAS Point Source Catalogue}

As before (Paper I), we have cross-correlated the entries in our deep optical
catalogue with the IRAS Point Source Catalogue (IRAS PSC). The IRAS PSC has been 
extensively used for studies of the large-scale structures in the Universe (\eg Saunders \etal 2000).
Our deep optical galaxy catalogue in the ZOA offers an opportunity to verify the 
performance of `blind' IRAS searches (\ie selecting galaxies purely on the basis
of their IRAS colours) close to the Galactic Plane, where source confusion could
play a significant role.

\subsection{IRAS PSC sources in the Zone of Avoidance}

We have selected galaxies in our optical galaxy catalogues which have a positional overlap
(within a radius of 2 arcmin) with sources in the IRAS PSC. Within this selection, there are
some galaxies for which the IRAS PSC association is unlikely because of their
unlikely colours or large positional offsets.  They are marked with `N' in
Table 1. They are not accepted as a galaxy with a credible IRAS PSC counterpart, nor are they listed
in Table 3. 

A sample page of the remaining galaxies is given in Table 3\footnote[1]{The IRAS cross-identifications 
in the Crux, respectively the GA region, are available in electronic format in their complete form at the CDS
via anonymous ftp to cdsarc.u-strasbg.fr (130.79.128.5) or via http://cdsweb.u-strasbg.fr/Abstract.html}.
The entries in Table 3 are:

\newcounter{fig2}
\begin{list} {\bf Column \arabic{fig2}:}{\usecounter{fig2} 
}

\item Identification in the IRAS Point Source Catalogue. If the IRAS name is followed by
a `Y', it is also in the IRAS galaxy list of Yamada \etal (1993), if followed by a `*', it
satisfies all the Yamada {\etal} selection criteria, but is not listed there.
 
\item Quality parameter of the IRAS PSC cross-identification. Depending on the probability that the 
cross-identification is correct, the following categories were defined: ({\bf I}) high certainty 
identification with IRAS PSC, ({\bf P})  possible match with IRAS PSC, but either the colour is 
atypical for galaxies or the separation with regard to the uncertainty ellipse relatively large,
({\bf Q}) questionable cross-identification because of large positional offset and/or unlikely IRAS-colour. 

\item The WKK identification number as in the optical galaxy catalogue (as given in Table 1). The `L' indicates
whether this is also a Lauberts (1982) galaxy.

\item As Column 4 in Table 1.
\item As Column 5 in Table 1.
\item As Column 8 in Table 1.
\item As Column 9 in Table 1.
\item As Column 13 in Table 1.
\item As Column 14 in Table 1.
\item As Column 16 in Table 1.

\item Angular separation in arcsec between the optical position (Columns 4 and 5) and the 
position quoted in the IRAS Point Source Catalogue.

\item The flux density at 12$\mu$m.
\item The flux density at 25$\mu$m.
\item The flux density at 60$\mu$m.
\item The flux density at 100$\mu$m.

\item The IRAS flux qualities at 12$\mu$m, 25$\mu$m, 60$\mu$m, and 100$\mu$m, where `1' 
indicates a lower limit, `2' an uncertain flux, and `3' a good quality flux.

\item The IRAS colour $col_1$ = $f_{12} \, f_{25} / (f_{60})^2$.
\item The IRAS colour $col_2$ = $f_{100}/f_{60}$.

\end{list}

\subsubsection{The Crux region}

211 Galaxies in our Crux catalogue have a positional overlap (within a radius of 2 arcmin) with
sources in the IRAS PSC. Of these galaxies, 64 are unlikely IRAS galaxies (N) because of their
atypical IRAS colours and large positional offset to the centre of the IRAS PSC source.

\subsubsection{The Great Attractor region}

Similarly, we found a positional overlap for 266 galaxies in the Great Attractor region.
Of these 266 galaxies, 71 galaxies have an unlikely counterpart.

\subsection{Properties of the IRAS PSC galaxies}

Only 171 of the 342 galaxies in the Crux and GA region listed in Table 3 have a certain IRAS counterpart, based on the second
and third colour criterion of Yamada \etal (1993), \ie $col_1 < 1$ and $0.8 < col_2 < 5.0$,
respectively (see Columns 17 and 18). They are labelled with an `{\bf I}'. Within 
this sample of 171 galaxies there are 10 IRAS PSC sources that have two galaxies associated with them, and one 
IRAS PSC source with three galaxies, indicated by `I2' and `I3', respectively, in Column 2. So there are 171 galaxies
associated with 159 IRAS PSC sources.

The number of galaxies with certain IRAS PSC cross-identification
is reduced to 165 (associated with 143 IRAS PSC sources) if a strict lower limit for the flux density at 
60$\mu$m, \ie $f_{60} = 0.6$ Jy, is imposed. 

Surprisingly, only 104 IRAS PSC sources were identified by Yamada \etal (1993), leaving 39 sources (= 27\%)
undetected. Note, that all the galaxies found by Yamada \etal (1993), that satisfy our selection
criterion of $D \ge 0\farcm2$, have been retrieved by our deep optical survey.

\begin{table*}[t]
 \normalsize
 \renewcommand{\baselinestretch}{0.65}
\caption{IRAS Galaxies in the Zone of Avoidance: The Crux Region}

\scriptsize

\begin{tabular*}{18cm}{
  p{15mm}  @{\extracolsep{0.5mm}} c @ {\extracolsep{1.5mm}} l @{\extracolsep{0mm}} 
% 1                        1                        2
% IRAS                     Y                        IR
  r  @{\extracolsep{1mm}} c @{\extracolsep{2mm}} 
% 3                       3                      
% RKK                     L                   
  l@{\extracolsep{2mm}} l @{\extracolsep{2mm}}
% 4                       5
% RA                      Dec
  r @{\extracolsep{0mm}}r @{\extracolsep{2mm}} 
% 6                     7                      
% gal l                 gal b                  
  p{6mm} @{\extracolsep{-0.5mm}} p{4.5mm} @{\extracolsep{0mm}}
% 8
% Dx                     d
  r @{\extracolsep{2mm}} 
%  9                   
%  Bj                   
  p{2.7mm} @{\extracolsep{-1.2mm}} p{2.7mm} @{\extracolsep{-1.2mm}} 
  p{2.7mm} @{\extracolsep{-1.2mm}} 
  p{2.7mm} @{\extracolsep{-0.5mm}} p{2.7mm} @{\extracolsep{0mm}} 
% 10a-e      
% T1-5
  r @{\extracolsep{0mm}} 
%  11
% Sep
r @{\extracolsep{-1mm}}r @{\extracolsep{-1mm}} 
r @{\extracolsep{-1mm}} r @{\extracolsep{4mm}} 
% 12 - 15
% IRAS fluxes at 12, 25,60,100
p{10mm} @{\extracolsep{0mm}}
%   16
%   Fluxqualities
  r @{\extracolsep{-1mm}} r @{\extracolsep{0mm}}
% 17                           18 
%col1                         col2
}
\hline 
\vspace{-1mm} \\
&&&
\multicolumn{14}{c}{optical} \vline &
\multicolumn{8}{c}{IRAS}\\
\vspace{-1mm} \\

\cline{5-17} \cline {18-25} \\
\vspace{-1mm} \\
\multicolumn{2}{l}{IRAS PSC} &
IR &
\ WKK&  & 
\ \ \ \ R.A. &
 \ \ \ Dec.& 
gal $\ell$ \ &
 gal $b \ $ & 
\multicolumn{2}{c}{$D \times d$}& 
${B}_{25}$ & 
\multicolumn{5}{c}{Type} &
\ Sep & 
\multicolumn{4}{c}{\ Flux Density} &  
\ \ Qual. &
\multicolumn{2}{c}{\ \ \ Color \ \ } \\
\vspace{-1mm} \\
\ \ \  Ident. & &  & & &(h\,\, m\,\, s) & ($\deg$\,\, $\arcmin$\,\,
 $\arcsec$) & ($\deg$) \ &($\deg$) \ 
%  & & (mm) & (mm) & \multicolumn{2}{c}{($\arcsec$)} & ($^{\rm m})$ & &
  & \multicolumn{2}{c}{($\arcsec$)} & ($^{\rm m}$) &
  \multicolumn{5}{c}{class.} &
 ( $\arcsec$) &
 $f_{12}$ & $f_{25}$ & $f_{60}$ & $f_{100}$ & &
 $col_1$ & $col_2$ \\
\vspace{-1mm} \\
  \multicolumn{2}{c}{(1)} & 
(2) &
  \multicolumn{2}{c}{\,\,\, (3)} & 
 \ \ \ \ (4) &
 \ \ \ \ (5) &
 (6) \ &
 (7) \  &
  \multicolumn{2}{c}{(8)} &
(9) &
\multicolumn{5}{c}{(10)} & 
\ (11) & 
\ \ \ \ (12) &
\ \ \ \ (13) &
\ \ \ \ (14) &
\ \ \ \ (15) &
\ \ (16) &
\ (17) &
(18) \\
\vspace{-1mm} \\
\hline 
\vspace{-1mm} \\
I10505-6906&  &  I  &  114&   & 10 50 35.6 & -69 06 32 & 292.73 & -8.90& \hfill 19x&\hfill 13 & 17.3 &  E& & & & &   4 &  0.31 &  0.25 &  0.59 &  1.15 & 1 1 3 1 &   0.22 &  1.95 \\
I10564-6923& Y&  I  &  150& L & 10 56 23.7 & -69 22 53 & 293.32 & -8.92& \hfill 77x&\hfill 31 & 15.5 &  S& & & & &  17 &  0.25 &  0.28 &  1.54 &  2.19 & 1 3 3 3 &   0.03 &  1.42 \\
I10565-6822& *&  I  &  151&   & 10 56 29.0 & -68 21 24 & 292.89 & -7.99& \hfill 19x&\hfill  9 & 18.5 &  I& & & &?&  91 &  0.36 &  0.25 &  0.65 &  1.94 & 1 1 1 3 &   0.21 &  2.98 \\
I10576-6926&  &  I  &  158& L & 10 57 35.4 & -69 26 32 & 293.44 & -8.93& \hfill 51x&\hfill 40 & 15.7 &  S& & &5& &  22 &  0.29 &  0.25 &  0.49 &  1.89 & 1 1 3 1 &   0.30 &  3.86 \\
I10593-6723& Y&  I  &  163&   & 10 59 20.8 & -67 23 10 & 292.73 & -6.99& \hfill 27x&\hfill 15 & 17.6 &   & & & & &   7 &  0.25 &  0.25 &  0.98 &  3.15 & 1 1 3 3 &   0.07 &  3.21 \\
\vspace{-1.3 mm} \\ 
I11073-6958& Y&  I  &  207&   & 11 07 23.1 & -69 59 00 & 294.45 & -9.08& \hfill 48x&\hfill 32 & 15.8 &  S&Y& &4& &  32 &  0.25 &  0.25 &  0.65 &  1.53 & 1 1 3 3 &   0.15 &  2.35 \\
I11083-6858& Y&  I  &  213& L & 11 08 22.9 & -68 58 11 & 294.14 & -8.11& \hfill 78x&\hfill 52 & 14.8 &  S&B& &3& &   6 &  0.25 &  0.25 &  0.68 &  1.96 & 1 1 3 3 &   0.14 &  2.88 \\
I11111-6859& Y&  I  &  224& L & 11 11 10.8 & -68 59 39 & 294.38 & -8.04& \hfill255x&\hfill 50 & 13.5 &  S& & &3& &  19 &  0.25 &  0.25 &  1.33 &  3.45 & 1 1 3 3 &   0.04 &  2.59 \\
I11173-6834&  &  I  &  247&   & 11 17 20.9 & -68 35 04 & 294.76 & -7.46& \hfill 34x&\hfill  8 & 18.1 &  S& & &6& &  45 &  0.35 &  0.27 &  0.40 &  1.86 & 3 1 1 1 &   0.59 &  4.65 \\
I11205-6909& Y&  I2 &  257&   & 11 20 31.1 & -69 08 21 & 295.22 & -7.88& \hfill 26x&\hfill 15 & 17.3 &  S& & & & &  92 &  0.25 &  0.25 &  1.75 &  4.70 & 1 1 3 3 &   0.02 &  2.69 \\
\vspace{-1.3 mm} \\ 
I11205-6909& Y&  I2 &  258&   & 11 20 32.9 & -69 10 05 & 295.23 & -7.91& \hfill 34x&\hfill 17 & 16.7 &  S& & &1& &  32 &  0.25 &  0.25 &  1.75 &  4.70 & 1 1 3 3 &   0.02 &  2.69 \\
I11245-6804&  &  P  &  269&   & 11 24 29.6 & -68 04 06 & 295.21 & -6.75& \hfill 40x&\hfill 28 & 15.8 &  S& & &1& &  20 &  0.65 &  0.25 &  0.39 &  2.72 & 1 1 3 1 &   1.07 &  6.97 \\
I11246-7123&  &  I  &  270&   & 11 24 35.1 & -71 25 13 & 296.31 & -9.92& \hfill 22x&\hfill 19 & 17.7 &  S& & &L& & 110 &  0.28 &  0.25 &  0.40 &  1.84 & 1 1 1 3 &   0.44 &  4.60 \\
I11256-7026& Y&  I  &  278&   & 11 25 41.6 & -70 27 07 & 296.08 & -8.98& \hfill 39x&\hfill 11 & 17.0 &  S& & &1& &   9 &  0.25 &  0.25 &  1.02 &  2.28 & 1 1 3 3 &   0.06 &  2.24 \\
I11281-6952&  &  I  &  285& L & 11 28 12.7 & -69 54 26 & 296.11 & -8.39& \hfill 54x&\hfill 28 & 16.2 &  S& & &7& &  99 &  0.28 &  0.25 &  0.40 &  1.18 & 1 1 1 3 &   0.44 &  2.95 \\
\vspace{-1.3 mm} \\ 
I11340-7031& Y&  I  &  301& L & 11 34  2.3 & -70 31 50 & 296.78 & -8.84& \hfill 69x&\hfill 38 & 15.5 &  S& & &3&:&   4 &  0.25 &  0.25 &  0.82 &  1.27 & 1 1 3 2 &   0.09 &  1.55 \\
I11366-6939&  &  Q  &  304& L & 11 36 39.1 & -69 41 05 & 296.75 & -7.96& \hfill 81x&\hfill 23 & 15.7 &  S& & &5& &  80 &  0.43 &  0.25 &  0.50 &  8.00 & 1 1 3 1 &   0.43 & 16.00 \\
I11470-6824& Y&  I  &  358&   & 11 47  2.2 & -68 24 55 & 297.32 & -6.50& \hfill 58x&\hfill  9 & 16.8 &  S& & &4& &   5 &  0.25 &  0.25 &  2.56 &  4.75 & 1 2 3 3 &   0.01 &  1.86 \\
I11485-7147&  &  I  &  370&   & 11 48 34.7 & -71 47 26 & 298.26 & -9.75& \hfill 35x&\hfill 24 & 16.7 &  S& & & & &   8 &  0.25 &  0.25 &  0.57 &  1.60 & 1 1 3 3 &   0.19 &  2.81 \\
I11508-5320& Y&  I  &  399&   & 11 50 53.9 & -53 20 11 & 294.30 &  8.28& \hfill 38x&\hfill 26 & 16.1 &  S& & &1& &   9 &  0.25 &  0.20 &  1.97 &  4.53 & 1 2 3 3 &   0.01 &  2.30 \\
\vspace{-1.3 mm} \\ 
I11512-5349&  &  I  &  404& L & 11 51 11.8 & -53 49 35 & 294.45 &  7.81& \hfill 65x&\hfill 27 & 15.9 &  S& & &2& &  11 &  0.44 &  0.26 &  0.52 &  2.22 & 1 1 3 2 &   0.42 &  4.27 \\
I11516-5353&  &  Q  &  411& L & 11 51 44.4 & -53 52 57 & 294.54 &  7.78& \hfill 60x&\hfill 40 & 15.6 &  S& &R& & & 100 &  3.67 &  1.23 &  0.40 &  4.75 & 3 3 1 1 &  28.21 & 11.88 \\
I11529-5638&  &  P  &  428&   & 11 52 53.8 & -56 38 16 & 295.31 &  5.12& \hfill 54x&\hfill 16 & 16.8 &   & & & & &   5 &  0.48 &  0.25 &  0.37 & 13.09 & 1 1 3 1 &   0.88 & 35.38 \\
I11550-5544&  &  Q  &  478&   & 11 55  8.8 & -55 42 47 & 295.42 &  6.09& \hfill 15x&\hfill  9 & 18.9 &  L& & & & & 103 &  4.10 &  1.85 &  0.49 &  8.96 & 3 3 1 1 &  31.59 & 18.29 \\
I11566-5307& Y&  I  &  499& L & 11 56 36.6 & -53 07 50 & 295.09 &  8.67& \hfill 87x&\hfill 56 & 14.5 &  F& & & & &   5 &  0.25 &  0.25 &  1.05 &  3.08 & 1 1 3 3 &   0.06 &  2.93 \\
\vspace{-1.3 mm} \\ 
I11579-5346& Y&  I  &  524&   & 11 57 59.5 & -53 46 31 & 295.43 &  8.08& \hfill 46x&\hfill 24 & 16.0 &  S& & &1& &  48 &  0.25 &  0.25 &  0.89 &  2.77 & 1 1 3 3 &   0.08 &  3.11 \\
I12002-5333& Y&  I  &  560&   & 12 00 13.2 & -53 33 26 & 295.71 &  8.35& \hfill 31x&\hfill 23 & 16.3 &  F& & & &?&   4 &  0.31 &  0.94 &  1.92 &  2.09 & 3 3 3 2 &   0.08 &  1.09 \\
I12018-5219&  &  I  &  584&   & 12 01 50.5 & -52 18 05 & 295.72 &  9.63& \hfill 32x&\hfill 32 & 16.6 &  S&Y& &5& &  77 &  0.25 &  0.25 &  0.44 &  1.46 & 1 1 1 3 &   0.32 &  3.32 \\
I12110-5602& Y&  I  &  768&   & 12 11  0.6 & -56 02 06 & 297.68 &  6.17& \hfill 22x&\hfill 16 & 17.0 &  L& & & & &  17 &  0.25 &  0.28 &  1.22 &  5.50 & 1 1 3 2 &   0.05 &  4.51 \\
I12113-5511& *&  I  &  774&   & 12 11 16.5 & -55 10 58 & 297.59 &  7.02& \hfill 35x&\hfill  9 & 17.2 &  S& & &M& &  89 &  1.09 &  0.36 &  0.72 &  1.88 & 3 3 3 1 &   0.76 &  2.61 \\
\vspace{-1.3 mm} \\ 
I12115-5252&  &  I  &  778&   & 12 11 27.9 & -52 51 56 & 297.28 &  9.32& \hfill 13x&\hfill 12 & 18.2 &   & & & & &  44 &  0.27 &  0.25 &  0.55 &  1.47 & 1 1 3 1 &   0.22 &  2.67 \\
I12116-6001& *&  I  &  784&   & 12 11 40.8 & -60 01 25 & 298.35 &  2.24& \hfill113x&\hfill 75 & 14.1 &  S& & & & &   6 &  3.23 &  8.24 & 89.96 &208.60 & 3 3 3 2 &   0.00 &  2.32 \\
I12116-5615& Y&  I  &  787&   & 12 11 42.6 & -56 15 53 & 297.81 &  5.96& \hfill 20x&\hfill 17 & 17.5 &  L& & & & &  11 &  0.36 &  1.10 &  9.68 & 12.34 & 3 3 3 3 &   0.00 &  1.27 \\
I12116-5809&  &  Q  &  788&   & 12 11 45.5 & -58 10 12 & 298.09 &  4.08& \hfill 59x&\hfill 11 & 17.0 &  S& & &5& &  77 &  0.34 &  0.25 &  0.72 & 27.64 & 3 1 1 1 &   0.16 & 38.39 \\
I12125-5746&  &  Q  &  801&   & 12 12 32.9 & -57 47 19 & 298.14 &  4.47& \hfill 19x&\hfill  8 & 18.3 &  S& & &2&:&  29 &  0.53 &  0.56 &  0.68 & 21.32 & 1 1 3 1 &   0.64 & 31.35 \\
\vspace{-1.3 mm} \\ 
I12150-5927&  &  P  &  860&   & 12 15 06.1 & -59 27 22 & 298.70 &  2.86& \hfill 22x&\hfill  9 & 18.0 &  S& & & & &  52 &  0.95 &  1.41 &  3.35 & 25.93 & 2 3 2 3 &   0.12 &  7.74 \\
I12154-5627& Y&  I  &  867&   & 12 15 28.2 & -56 27 46 & 298.36 &  5.84& \hfill 47x&\hfill 35 & 16.0 &  L& & & & &   7 &  0.25 &  0.25 &  1.05 &  3.81 & 1 1 3 3 &   0.06 &  3.63 \\
I12181-7108& Y&  I  &  915&   & 12 18 10.3 & -71 08 48 & 300.49 & -8.70& \hfill 16x&\hfill 11 & 18.6 &   & & & & &   5 &  0.33 &  0.25 &  0.74 &  2.11 & 1 1 3 1 &   0.15 &  2.85 \\
I12195-5436&  &  I  &  945&   & 12 19 31.8 & -54 36 39 & 298.70 &  7.74& \hfill 28x&\hfill  7 & 17.7 &  S& & &L&?&  53 &  0.33 &  0.28 &  0.40 &  1.48 & 3 1 1 1 &   0.58 &  3.70 \\
I12196-5315&  &  I  &  950&   & 12 19 43.3 & -53 15 36 & 298.57 &  9.09& \hfill 15x&\hfill  9 & 18.5 &  I& & & & &  32 &  0.28 &  0.35 &  0.50 &  1.21 & 1 1 1 3 &   0.39 &  2.42 \\
\vspace{-1.3 mm} \\ 
I12199-5820& Y&  I  &  957& L & 12 19 54.4 & -58 20 20 & 299.18 &  4.05& \hfill101x&\hfill 27 & 15.0 &  S& & &2&:&  31 &  0.25 &  0.25 &  2.37 &  5.63 & 1 1 3 3 &   0.01 &  2.38 \\
I12200-5404&  &  P  &  959&   & 12 19 59.5 & -54 04 38 & 298.71 &  8.28& \hfill 50x&\hfill 19 & 15.8 &  S& & &M& &  43 &  0.25 &  0.33 &  0.54 &  3.21 & 1 1 3 1 &   0.28 &  5.94 \\
I12203-5456&  &  I  &  968&   & 12 20 21.1 & -54 56 38 & 298.86 &  7.43& \hfill 17x&\hfill 12 & 17.4 &  E& & & & &  13 &  0.37 &  0.25 &  0.50 &  1.47 & 1 1 3 1 &   0.37 &  2.94 \\
I12265-7203& *&  I  & 1072&   & 12 26 30.6 & -72 03 10 & 301.25 & -9.53& \hfill 23x&\hfill 15 & 17.3 &  S& & &1&:&  25 &  0.26 &  0.25 &  0.60 &  2.44 & 1 1 3 1 &   0.18 &  4.07 \\
I12269-5404& Y&  I  & 1078& L & 12 27 02.7 & -54 05 14 & 299.75 &  8.37& \hfill 74x&\hfill 16 & 15.9 &  S& & &1& &  48 &  0.33 &  0.25 &  0.99 &  2.12 & 1 1 3 2 &   0.08 &  2.14 \\
\vspace{-1.3 mm} \\ 
I12282-5451&  &  P  & 1089& L & 12 28 16.2 & -54 51 40 & 299.99 &  7.62& \hfill 85x&\hfill 54 & 14.5 &  I& & & & &  13 &  0.29 &  0.25 &  0.71 &  6.23 & 1 1 3 1 &   0.14 &  8.77 \\
I12295-6803& Y&  I  & 1110& L & 12 29 30.2 & -68 03 38 & 301.17 & -5.53& \hfill101x&\hfill 65 & 14.6 &  S&B& &5& &   4 &  0.44 &  0.25 &  0.90 &  3.81 & 1 1 3 1 &   0.14 &  4.23 \\
I12369-5454&  &  I  & 1233&   & 12 36 56.3 & -54 54 54 & 301.25 &  7.64& \hfill 31x&\hfill 20 & 16.9 &  S&B& &5& &   9 &  0.41 &  0.25 &  0.60 &  2.72 & 1 1 3 1 &   0.28 &  4.53 \\
I12400-5843& Y&  I  & 1291&   & 12 40 04.2 & -58 43 38 & 301.84 &  3.85& \hfill 26x&\hfill  9 & 17.9 &  S& & & & &  14 &  0.25 &  0.30 &  2.06 &  6.10 & 1 3 3 3 &   0.02 &  2.96 \\
I12446-5340& Y&  I  & 1373& L & 12 44 41.2 & -53 40 34 & 302.35 &  8.92& \hfill110x&\hfill 78 & 14.1 &  I& & & &?&  25 &  0.28 &  0.25 &  0.78 &  2.04 & 1 1 3 1 &   0.12 &  2.62 \\
\vspace{-1.3 mm} \\ 
I12447-5316& Y&  I  & 1379& L & 12 44 46.3 & -53 16 41 & 302.36 &  9.32& \hfill 58x&\hfill 46 & 14.8 &  E& & & &?&   5 &  0.26 &  0.26 &  2.38 &  3.99 & 1 2 3 3 &   0.01 &  1.68 \\
I12483-5316& Y&  I2 & 1434&   & 12 48 21.5 & -53 16 32 & 302.90 &  9.32& \hfill 20x&\hfill  7 & 18.0 &  L& & & &?&  20 &  0.41 &  0.25 &  1.33 &  1.75 & 1 1 3 1 &   0.06 &  1.32 \\
I12483-5316& Y&  I2 & 1435&   & 12 48 21.7 & -53 16 54 & 302.90 &  9.32& \hfill 19x&\hfill 15 & 17.3 &  S& & & & &   4 &  0.41 &  0.25 &  1.33 &  1.75 & 1 1 3 1 &   0.06 &  1.32 \\
I12547-5307& Y&  I  & 1576& L & 12 54 44.8 & -53 07 19 & 303.87 &  9.47& \hfill 94x&\hfill 54 & 15.1 &  S& & &5& &   5 &  0.25 &  0.25 &  1.10 &  2.75 & 1 1 3 3 &   0.05 &  2.50 \\
I12571-5543&  &  P  & 1642&   & 12 57 07.9 & -55 43 01 & 304.15 &  6.86& \hfill 34x&\hfill 23 & 16.2 &  F& & & & &   4 &  0.30 &  0.25 &  0.64 & 12.77 & 1 1 3 1 &   0.18 & 19.95 \\
\vspace{-1.3 mm} \\ 
I12584-7101& *&  I  & 1674&   & 12 58 24.7 & -71 01 44 & 303.77 & -8.44& \hfill 24x&\hfill 12 & 18.1 &  S& & & & &  26 &  0.30 &  0.25 &  1.05 &  2.09 & 1 1 3 2 &   0.07 &  1.99 \\
I12585-5300& Y&  I  & 1680&   & 12 58 37.0 & -53 00 19 & 304.47 &  9.57& \hfill 31x&\hfill 31 & 16.5 &  S& & &5&:&  21 &  0.29 &  0.25 &  0.86 &  1.87 & 1 1 3 1 &   0.10 &  2.17 \\
I12588-5349& Y&  I  & 1685&   & 12 58 50.0 & -53 49 23 & 304.47 &  8.75& \hfill 22x&\hfill 16 & 18.0 &  S& & & &?&  13 &  0.25 &  0.41 &  2.36 &  2.70 & 1 3 3 3 &   0.02 &  1.14 \\
I12593-5601&  &  P  & 1694& L & 12 59 23.0 & -56 01 28 & 304.46 &  6.55& \hfill 78x&\hfill 17 & 16.1 &  S& & &4&:&  10 &  0.45 &  0.40 &  0.60 & 11.59 & 1 1 3 1 &   0.50 & 19.32 \\
I12594-5552& *&  I  & 1696&   & 12 59 27.0 & -55 52 48 & 304.48 &  6.69& \hfill 78x&\hfill 15 & 16.7 &  S& & &5& &   6 &  0.28 &  0.25 &  0.88 &  3.48 & 1 1 3 3 &   0.09 &  3.95 \\
\vspace{-1.3 mm} \\ 
I12595-5403& Y&  I  & 1698&   & 12 59 29.3 & -54 03 40 & 304.56 &  8.51& \hfill 17x&\hfill 12 & 17.6 &  L& & & & &  15 &  0.25 &  0.25 &  1.65 &  2.78 & 1 1 3 1 &   0.02 &  1.68 \\
I12597-6547&  &  P  & 1707&   & 12 59 44.3 & -65 47 43 & 304.10 & -3.22& \hfill 26x&\hfill 11 & 17.5 &  S& & & & &   4 &  0.25 &  0.45 &  4.07 & 48.19 & 1 3 3 1 &   0.01 & 11.84 \\
I13003-7151& Y&  I  & 1716&   & 13 00 21.9 & -71 51 16 & 303.90 & -9.27& \hfill 51x&\hfill 16 & 16.9 &  S& & &3& &   6 &  0.25 &  0.25 &  0.63 &  1.62 & 1 1 3 3 &   0.16 &  2.57 \\
I13005-5231&  &  I  & 1719&   & 13 00 33.3 & -52 31 09 & 304.78 & 10.04& \hfill 31x&\hfill 12 & 17.3 &  S& & &2& &  39 &  0.40 &  0.26 &  0.49 &  2.19 & 1 1 3 1 &   0.43 &  4.47 \\
I13009-5608&  &  Q  & 1734&   & 13 00 58.0 & -56 08 24 & 304.68 &  6.42& \hfill 28x&\hfill 15 & 17.1 &  S& & &M& &   2 &  0.42 &  0.25 &  1.00 & 10.27 & 1 1 3 1 &   0.11 & 10.27 \\
\vspace{-1.3 mm} \\ 
I13031-5717& Y&  I2 & 1783&   & 13 03 08.6 & -57 17 54 & 304.92 &  5.25& \hfill 19x&\hfill  7 & 18.1 &  S& & &0& &  32 &  0.25 &  0.42 &  2.92 &  5.16 & 1 3 3 3 &   0.01 &  1.77 \\
I13031-5717& Y&  I2 & 1784& L & 13 03 10.0 & -57 17 20 & 304.92 &  5.26& \hfill 70x&\hfill 55 & 14.9 &  S& & &5& &  32 &  0.25 &  0.42 &  2.92 &  5.16 & 1 3 3 3 &   0.01 &  1.77 \\
I13042-5351& Y&  I  & 1822& L & 13 04 18.1 & -53 51 34 & 305.28 &  8.67& \hfill 60x&\hfill 30 & 15.9 &  S& &R&5& &   9 &  0.36 &  0.25 &  0.68 &  3.17 & 1 1 3 1 &   0.19 &  4.66 \\
I13051-5358&  &  P  & 1851& L & 13 05 12.3 & -53 58 40 & 305.41 &  8.55& \hfill 59x&\hfill 42 & 15.4 &  S&Y& &5& &  10 &  0.25 &  0.25 &  0.51 &  3.24 & 1 1 3 1 &   0.24 &  6.35 \\
I13052-5711& Y&  I  & 1853&   & 13 05 15.0 & -57 11 31 & 305.21 &  5.34& \hfill 32x&\hfill 32 & 16.3 &  S& & &9&?&   5 &  0.25 &  0.44 &  7.82 & 11.73 & 1 3 3 3 &   0.00 &  1.50 \\
\vspace{-1.3 mm} \\ 
I13064-5712& Y&  I  & 1883&   & 13 06 26.2 & -57 12 17 & 305.37 &  5.32& \hfill 32x&\hfill 12 & 17.3 &  S& & &3& &  20 &  0.27 &  0.25 &  0.68 &  3.34 & 1 1 3 1 &   0.15 &  4.91 \\
I13073-5342&  &  Q  & 1906&   & 13 07 23.2 & -53 42 17 & 305.75 &  8.80& \hfill 13x&\hfill 13 & 18.0 &  L& & & & &  14 &  0.27 &  0.35 &  0.69 &  8.43 & 1 1 3 1 &   0.20 & 12.22 \\
I13081-5304& *&  I  & 1935&   & 13 08 09.7 & -53 04 25 & 305.92 &  9.42& \hfill 27x&\hfill  9 & 18.4 &  S& & &L& &   8 &  0.25 &  0.28 &  1.88 &  9.10 & 1 2 3 1 &   0.02 &  4.84 \\
I13085-5807&  &  Q  & 1946&   & 13 08 34.8 & -58 07 04 & 305.59 &  4.38& \hfill 20x&\hfill  8 & 18.4 &  S& & &L& &  10 &  0.35 &  0.25 &  0.87 & 22.05 & 1 1 3 1 &   0.12 & 25.34 \\
I13089-6518& *&  I  & 1957&   & 13 08 57.5 & -65 18 28 & 305.08 & -2.79& \hfill 19x&\hfill 13 & 17.6 &   & & & & &   1 &  0.58 &  2.77 & 24.02 & 27.42 & 3 3 3 3 &   0.00 &  1.14 \\
\vspace{-1.3 mm} \\ 
\hline
 \end{tabular*}
 \normalsize
\label{iras}
\end{table*}
\addtocounter{table}{-1}
\clearpage
%\newpage

\begin{figure}
 \resizebox{\hsize}{!}{\includegraphics{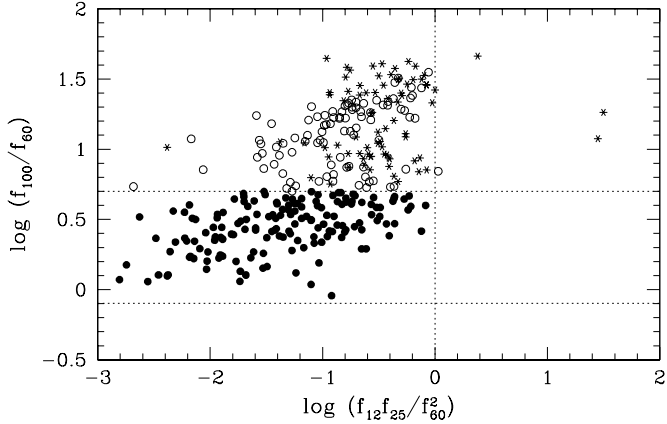}}
\caption{IRAS colour-colour diagram ($f_{12}f_{25}/f^2_{60}$ vs.~$f_{100}/f_{60}$)
for galaxies in the Crux and Great Attractor regions that have a certain IRAS counterpart (filled circles),
a possible IRAS counterpart (open circles) and a questionable IRAS association (stars).}
\label{irascol}
\end{figure}

A further 94 galaxies have a likely IRAS counterpart, primarily based on their positional overlap.
These galaxies are indicated by a `{\bf P}' in Column 2. In this sample there are
two IRAS PSC sources that have two galaxies belonging to them, indicated by `P2' in Column 2.
72 IRAS PSC sources of this sample have $f_{60} \ge 0.6$ Jy. 
These galaxies do not comply with the preset colour restriction of $f_{100}/f_{60} \le 5$ (see 
Fig.~\ref{irascol}), but they are all bona fide galaxies. 

Finally, there are 77 galaxies for which their association with an IRAS PSC source is questionable,
mostly because of their large positional offset to the IRAS sources and sometimes because of their
unlikely IRAS colours (see Fig.~\ref{irascol}). These galaxies are indicated by a `{\bf Q}'. 
The average properties of galaxies in each of these samples are listed in Table 4.

From Table 4 it is clear that the IRAS PSC traces a population of large 
($<D> = 62''$) and bright ($<B_{25}> = 16\fm1$) galaxies. They are predominantly spiral
galaxies ($\sim$80\%). The `P' sample consists of large and bright galaxies, as bright
as the certain (I) IRAS PSC galaxies.

\addtocounter{table}{+1}
\begin{table}[h]
 \caption{Properties of galaxies with an IRAS PSC association.}
 \label{irastab}
\begin{center}
\begin{tabular}{ccccr}
\hline
\vspace{-2.3mm} \\
Sample &  $<B_{25}>$ & $<D>$  & Separation & $N_{gal}$ \\
\vspace{-1.3mm} \\
\hline
\vspace{-1.3mm} \\
 {\bf I}  & \hfill  $16\fm1$ & \hfill   $62\arcsec$ & $27\arcsec \pm 28\arcsec$ & 159 \\
 {\bf P}  & \hfill  $16\fm1$ & \hfill   $54\arcsec$ & $27\arcsec \pm 27\arcsec$ & 92 \\
 {\bf Q}  & \hfill  $17\fm5$ & \hfill   $27\arcsec$ & $69\arcsec \pm 32\arcsec$ & 72 \\
\vspace{-1.3mm} \\
\hline
\end{tabular}
\end{center}
\end{table}

In Fig.~\ref{irascol} we show a colour-colour diagram ($f_{12}f_{25}/f^2_{60}$ vs.~$f_{100}/f_{60}$)
of the galaxies with an IRAS PSC assocations. The galaxies with a certain IRAS PSC counterpart (I) are
shown as filled circles, the possible counterparts (P) as open circles and the questionable assocations
(Q) as stars. The selection criteria of Yamada \etal (1993) are indicated by the dashed lines.
Yamada (1994) already noted that the upper limit in the $f_{100}/f_{60}$ colour restriction, imposed to 
limit the contamination by Galactic cirrus, will make the `blind' IRAS search less complete. 
As we see here, this incompleteness is substantial ($\sim$37\%).

We have obtained redshifts for 35 of the galaxies in the `P' sample (38\%) (Woudt 1998) and 77\% 
of them lie in the redshift range 1400 -- 5500 km s$^{-1}$ and are hence important tracers of the large-scale
structures in the nearby Universe. The remaining 23\% (8 galaxies, all in the GA region) are located beyond 
10\,000 km s$^{-1}$. All these galaxies are, however, missed in blind IRAS searches.

\begin{figure}
 \resizebox{\hsize}{!}{\includegraphics{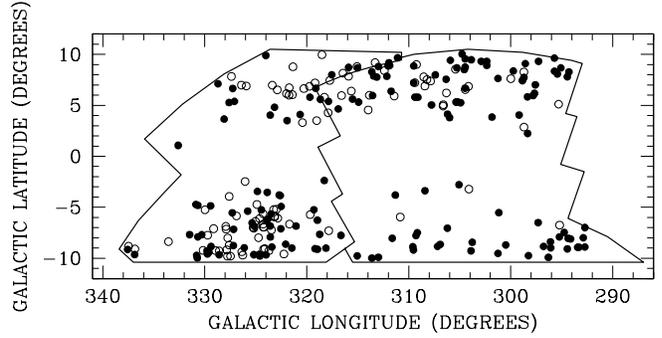}}
\caption{The distribution in Galactic coordinates of the 251 galaxies in the Crux and GA regions
with either a reliable IRAS PSC cross-identification (filled circles, 159 galaxies) or a possible
IRAS PSC cross-identification (open circles, 92 galaxies).}
\label{irasdis}
\end{figure}

\subsection{The distribution of the IRAS PSC galaxies}

\subsubsection{The Crux region}

Fig.~\ref{irasdis} shows the distribution (in Galactic coordinates) of the galaxies in the Crux region with an
IRAS PSC association. The galaxies in the `{I}' sample are shown as filled circles and the galaxies
in the `{P}' sample as open circles. The majority of the galaxies are located north of the Galactic
Plane, in particular the galaxies in the `{P}' sample; 26 galaxies (out of 29) are located at positive
Galactic latitudes and these galaxies are most likely associated with the Norma 
supercluster (Fairall \etal 1998).

The overdensity at $(\ell, b) = (315\deg, -8\deg)$ noted in Fig.~\ref{cruxgaf2} is totally absent in Fig.~\ref{irasdis}.
There are, in fact, no galaxies at all in the IRAS PSC in this part of the sky. This again suggests that
this overdensity is more distant. 

\subsubsection{The Great Attractor region}

The distribution of the galaxies in the GA region with a certain and probable IRAS counterpart is
also shown in Fig.~\ref{irasdis}. Most of the galaxies are located below the Galactic Plane
in an extended region around the Norma cluster. The IRAS PSC galaxies trace the general
overdensity of galaxies in the GA region. They do not, however, reveal the
rich and nearby Norma cluster in the way our deep optical search has done.

\begin{figure}
 \resizebox{\hsize}{!}{\includegraphics{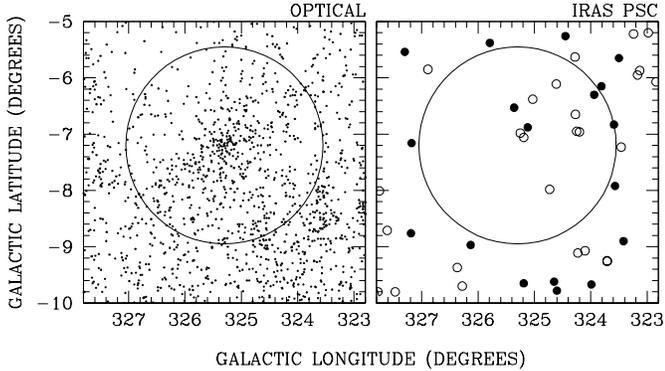}}
\caption{The distribution in Galactic coordinates of the galaxies from our deep optical galaxy
catalogue centred on the Norma cluster (left panel), and the galaxies with an IRAS PSC 
counterpart in the same region (right panel, symbols as in Fig.~\ref{irasdis}).}
\label{gafig9}
\end{figure}

In Fig.~\ref{gafig9} we show a comparison of the galaxy distribtion in the optical 
and in the far infrared around the rich galaxy cluster of Norma. Unlike the centrally
condensed overdensity in the optical distribution (left panel of Fig.~\ref{gafig9}), 
the IRAS PSC data merely show an extended overdensity (see Fig.~\ref{irasdis}) with no clear
peak around the central part of the Norma cluster (right panel of Fig.~\ref{gafig9}).
This is in line with the expectation that the IRAS data do trace the general large-scale
structures, but are insensitive to the elliptical galaxies that reside in rich clusters.

\section{Remarks on the Great Attractor}

Our deep optical search for partially obscured galaxies in the Crux and Great Attractor
region has revealed several overdensities which belong to the Great Attractor. These surveys
show that a fair fraction of the mass of the Great Attractor was indeed previously 
hidden behind the Milky Way. In this section, we will summarise our results of the
Great Attractor and discuss the possibility that other rich clusters of galaxies might still be
hidden behind the remaining optical Zone of Avoidance ($A_B \ge 3^{\rm m}$).

\subsection{Unveiled structures}

Following our galaxy search, the below listed structures unveiled (and recognised)
at low Galactic latitude can be associated with the Great Attractor (see also Fig.~\ref{cruxf1new}):

\begin{itemize}
\item[$\bullet$]{A broad, extended overdensity north of the Galactic Plane at 
$305\deg \le \ell \le 320\deg$.}
\item[$\bullet$]{The low-mass Centaurus--Crux galaxy cluster at $(\ell, b, v) \approx (305.5\deg, +5.5\deg,
6214$ km s$^{-1}$) (Fairall \etal 1998).}
\item[$\bullet$]{The rich and nearby Norma cluster at $(\ell, b, v) \approx (325.3\deg, -7.2\deg, 4844$~km~s$^{-1}$) 
(Kraan-Korteweg~et~al. 1996; Woudt \etal 1999).}
\end{itemize}

Tentative evidence from the magnitude distribution of the galaxies in the GA region, 
that the Great Attractor survey region is dominated by a rich cluster at the distance
of the Great Attractor (see Fig.~\ref{gafig5b} and the discussion in Sect.~6.4) is 
confirmed beyond doubt by our follow-up redshift surveys (Kraan-Korteweg \etal 1996; Woudt \etal 1999).
With a velocity dispersion of 896 km s$^{-1}$, the Norma cluster has a mass of the same order as the Coma cluster.
This is confirmed by ROSAT PSPC observations of this cluster (B\"ohringer \etal 1996).
Simulations have furthermore shown that the well-known Coma cluster would appear the same as the Norma cluster
if Coma were located behind the Milky Way at the location of the Norma
cluster (Woudt 1998). 

The Norma cluster is the most massive cluster known to date in the Great Attractor overdensity. Its optical
appearance, however, is not very prominent, due to the obscuring effects of the Galactic foreground extinction. It is
the most likely candidate to mark the Great Attractor's hitherto unseen core (Kraan-Korteweg \etal 1996).

\subsubsection{The Norma cluster}

Irrespective of its position within the Great Attractor overdensity, the Norma cluster is an important cluster
in its own right. The cluster is comparable to the Coma cluster, but is located closer (Woudt 1998), \ie
it is the nearest rich cluster in the local Universe. 
It is X-ray bright (B\"ohringer \etal 1996), H\,I deficient (Vollmer \etal 2001), reveals
signs of an ongoing merger close to the centre of the cluster (B\"ohringer \etal 1996; Woudt 1998), harbours an
infalling spiral-rich subgroup at a distance of 2--3 $h_{50}^{-1}$ Mpc from the centre (Woudt 1998), contains
2 central cD galaxies (like the Coma cluster) and has two strong radio continuum sources, namely PKS1610-608 (the 
central Wide-Angle-Tail source) and PKS1610-605 (an extended Head-Tail source) (Jones \& McAdam 1992). 
Due to the proximity of the Norma cluster, it is a good laboratory to study the interaction of galaxies in a rich cluster
with the Intracluster Medium.

Within the Abell radius of the Norma cluster there are 603 galaxies in our ZOA catalogue. We have obtained
redshifts for 266 galaxies, 219 of which are bona fide cluster members. More redshifts were obtained
in 2001 with 2dF observations of this cluster. A full dynamical analysis of the Norma cluster will 
be presented elsewhere (Woudt \etal in prep.). We have furthermore obtained deep $R$-band images with the
ESO/MPG 2.2-m telescope (with the Wide Field Imager) covering the entire Abell radius of the Norma cluster.
These data will allow a good determination of the luminosity function of this nearby and rich cluster.
In combination with recently obtained pointed $K'$ observations for $\sim$50 elliptical galaxies in the Norma cluster,
a distance to the Norma cluster can be derived through a Fundamental Plane analysis. The observed $(R-K')$ colours
of the elliptical galaxies in this sample will provide additional information on 
the Galactic foreground extinction (see discussion in Sect.~5.2). 

A reliable distance to the Norma cluster (with an uncertainty of $\la 100$ km s$^{-1}$) is essential 
in distinguishing between the various existing models of the GA; a nearby GA (Tonry \etal 2000), a more distant
GA (Kolatt \etal 1995), or a GA which partakes in a cosmic flow to even larger distances.

\subsubsection{The galaxy distribution}

\begin{figure*}
 \resizebox{\hsize}{!}{\includegraphics{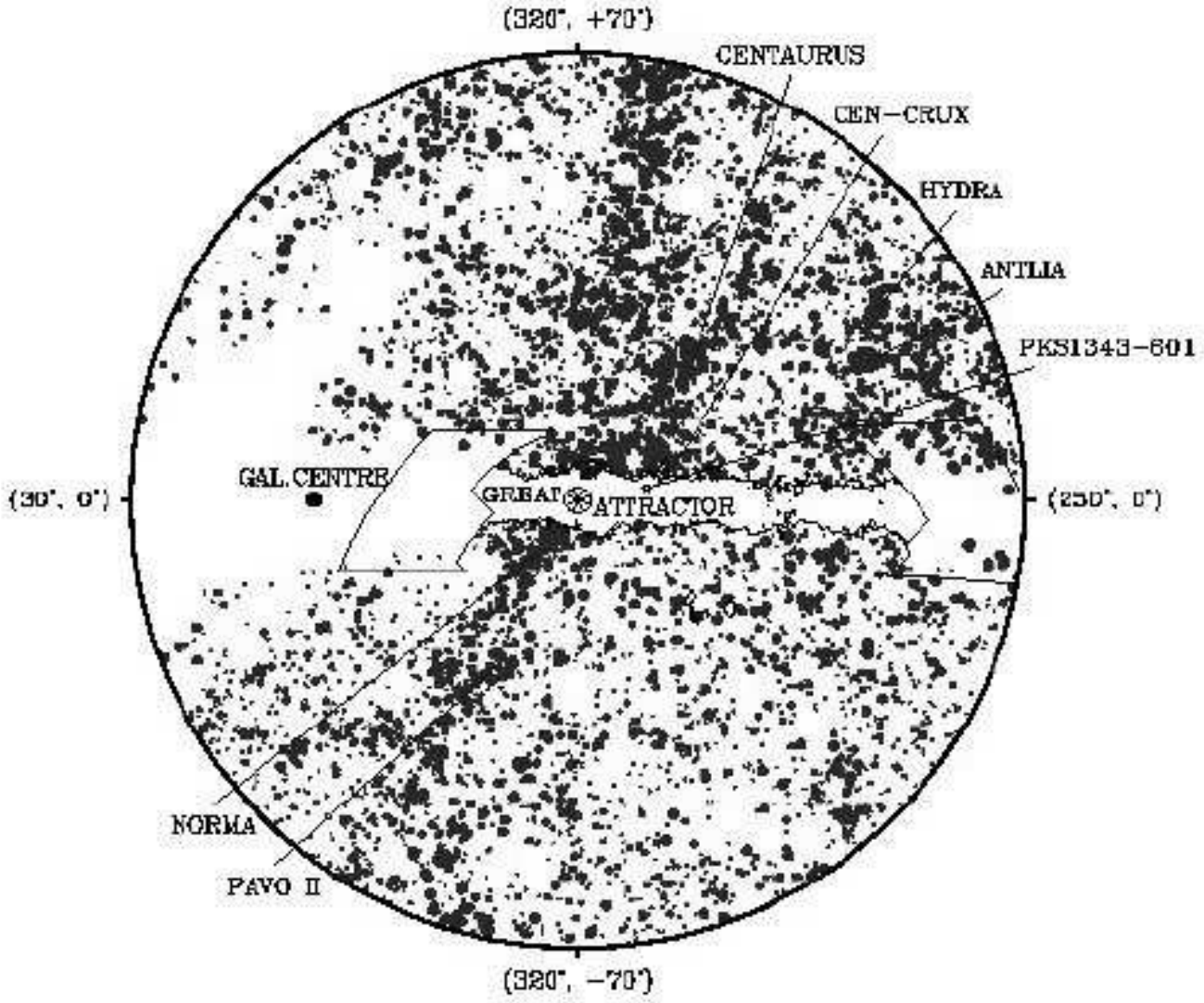}}
\caption{An equal area projection of galaxies with $D^0 \ge 1\farcm3$ and $A_B \le 3^{\rm m}$ centred
on the Great Attractor at $(\ell, b) = (320\deg, 0\deg)$ within a radius of 70$\deg$. The galaxies are taken from the ESO Uppsala Catalogue
(Lauberts 1982), the Uppsala General Catalogue UGC (Nilson 1973), the Morphological Catalogue of Galaxies MCG
(Vorontsov-Velyaminov \& Archipova 1963-74), {\sl and} our Zone of Avoidance catalogues which significantly reduce the optical ZOA.
Over the extent of our survey region, we have marked the $A_B = 3^{\rm m}$ extinction-contour. Our
search areas in progress (the Scorpius region to the left, and the Vela region to the right) are indicated by the solid line.
Prominent overdensities are marked and labelled. The galaxies are diameter-coded as in Fig.~\ref{cruxf1}.
}
\label{cruxf1new}
\end{figure*}

Fig.~\ref{cruxf1new} shows the most complete view of the optical galaxy distribution
in the Great Attractor region to date. This equal area projection is centred on the Great Attractor
at $(\ell, b) = (320\deg, 0\deg)$. All galaxies with extinction-corrected diameters larger than
$D^0 \ge 1\farcm3$ limited by $A_B \le 3^{\rm m}$ are shown. Galaxies were taken 
from the Lauberts (1982) catalogue, the Uppsala General Catalogue UGC (Nilson 1973), the Morphological 
Catalogue of Galaxies MGC (Vorontsov-Velyaminov \& Archipova 1963-74), and our Zone of Avoidance Catalogues 
(Paper I, and this paper). Details on diameter corrections for galaxies in the Lauberts, 
UGC and MCG catalogues are given by Kraan-Korteweg (2000b).

The resulting galaxy distribution shows a strong concentration of galaxies towards the Galactic Plane
on either side of the centre of the Great Attractor.
This concentration of galaxies, previously unnoticed (compare with Fig.~\ref{cruxf1}), most likely is part of 
the Centaurus Wall and the Norma Supercluster (see also Fairall \etal 1998) and hence associated with the
Great Attractor.
Comparing Fig.~\ref{cruxf1new} with Fig.~1 of Paper I, shows that we have greatly
reduced the optical Zone of Avoidance (by more than 50\%) and have unveiled a large fraction of the 
Great Attractor overdensity, previously unseen, but also that the Great Attractor region is not 
entirely unveiled yet.

The remaining ZOA ($A_B \ge 3^{\rm m}$) is best surveyed at higher wavelengths. This is done already in the near infrared 
by DENIS (Schr\"oder \etal 1999) and 2MASS (Jarrett \etal 2000), at 21 cm by the Parkes HI ZOA Sky Survey 
(Juraszek \etal 2000; Henning \etal 2000) and in the X-rays (B\"ohringer \etal 2000; Ebeling \etal 2000).

\subsection{What remains hidden?}

One can, for instance, not exclude that another Norma-like cluster has remained hidden behind the
reduced optical Zone of Avoidance, \ie at $A_B \ge 3^{\rm m}$. The ZOA survey greatly enhanced the 
optical appearance of the Norma cluster within the extinction interval $A_B = 1^{\rm m} - 2^{\rm m}$. 
At $A_B \ge 3^{\rm m}$ we become increasingly incomplete, however, and at $A_B \ge 5^{\rm m}$ we stop
finding galaxies altogether in the optical. Even a rich, nearby cluster cannot be detected
by optical surveys at the highest extinction levels. 

Rich clusters can, however, be detected through a fair bit of extinction by X-ray surveys (B\"ohringer \etal 2000,
Ebeling \etal 2000) although even these searches are limited by the current survey material. Alternatively,
a strong central radio source, such as PKS1610-608 in the Norma cluster, could point to unidentified clusters.

Exactly such a source lies in the deepest layers of the Galactic foreground extinction ($A_B = 12^{\rm m}$)
at $(\ell, b, v) = (309.7\deg, +1.7\deg, 3872$ km s$^{-1}$), {\sl right in the Great Attractor overdensity}
(Kraan-Korteweg \& Woudt 1999).
We are currently involved in a deep infrared survey ($J, H, K\arcmin$) with the 1.4-m Infrared Survey Facility
of Nagoya University at the South African Astronomical Observatory, to see if this strong radio-source (PKS1343-601)
is the central source of a highly obscured, rich cluster (Nagayama {\etal}, in prep.). 
If there is indeed another rich cluster in the Great Attractor
region, this would have serious implications for our understanding of the formation of this nearby
massive overdensity.

\begin{acknowledgements}
Part of this survey was performed at the Kapteyn Astronomical Institute of the University 
of Groningen. Their support is greatfully acknowledged. PAW is supported by strategic funds made
available by the University of Cape Town. He further acknowledges financial support from
the National Research Foundation. RCKK thanks CONACyT for their support (research grant
27602E). The authors acknowledge the Referee, Dr.~R.~Peletier, for the useful comments and suggestions.
This research has made use of 
the NASA/IPAC Extragalactic Database (NED), which is operated by the Jet 
Propulsion Laboratory, Caltech, under contract with the National Aeronautics 
and Space Administration.

\end{acknowledgements}

\end{document}